\documentclass[floatfix,superscriptaddress,amsmath,aps,prb]{revtex4}

\pdfoutput=1

\usepackage{t1enc}
\usepackage[final]{graphicx}
\usepackage{epsfig}
\usepackage{bm}
\usepackage[normalem]{ulem}
\usepackage[usenames]{xcolor}
\usepackage{times}
\usepackage{verbatim}
\usepackage{amssymb}
\usepackage{comment}
\usepackage{adjustbox}
\graphicspath{{images/}}

\usepackage{lipsum}
\usepackage{xfrac}

\makeatletter\begin{document}

\title{Melting of the two-dimensional solid phase in the Gaussian-core model}

\author{Alejandro Mendoza-Coto}
\email{alejandro.mendoza@ufsc.br}
\affiliation{Departamento de Física, Universidade Federal de Santa Catarina, 88040-900 Florianópolis, Brazil}
\affiliation{Max-Planck-Institut für Physik komplexer Systeme, Nöthnitzer Str. 38, 01187 Dresden, Germany}
\author{Valéria Mattiello}
\affiliation{Departamento de Física, Universidade Federal de Santa Catarina, 88040-900 Florianópolis, Brazil}
\author{Rômulo Cenci}
\affiliation{Departamento de Física, Universidade Federal de Santa Catarina, 88040-900 Florianópolis, Brazil}
\author{Nicolò Defenu}
\affiliation{Institut für Theoretische Physik, Eidgeössische Technische Hochschule Zürich, 8093 Zürich, Switzerland}
\author{Lucas Nicolao}
\affiliation{Departamento de Física, Universidade Federal de Santa Catarina, 88040-900 Florianópolis, Brazil}
\date{\today}

\begin{abstract}

A general theory for the melting of two dimensional solids explaining the universal and non-universal properties is an open problem up to date. Although the celebrated KTHNY theory have been able to predict the critical properties of the melting transition in a variety cases, it is already known that it is not able to capture the occurrence of first order transitions observed in certain systems as well as it doesn't provide a clear way to calculate the melting temperature for a specific model.
In the present work we have developed an analytical method that combines Self Consistent Variational Approximation with the Renormalization Group in order to deal simultaneously with the phonon fluctuations and the topological defects present in the melting process of two dimensional crystals. The method was applied with impressive success to the study of the phase diagram of the Gaussian-core model, capturing not only the reentrant feature of its 2D solid phase, but also the related critical temperatures as a function of the density in quantitative detail. The developed method can be directly applied to study the melting of any hexagonal simple crystal formed by particles interacting through any finite pairwise interaction potential. Additionally, it has the potential to explain the occurrence of first order transitions in the melting process of two dimensional crystals.

\end{abstract}

\maketitle

\section{Introduction}

Two dimensional systems described by continuous microscopic variables present unusual phase transitions between the low and high temperature phases. This is related with the fact that in two dimensions thermal fluctuations are quite strong, preventing the stabilization of a long range ordered phase \cite{Peierls1935, Landau1937, MeWa1966,Ho1967}. In this scenario, a vestigial quasi long-range order is often observed at low temperatures with slowly power law decaying correlations functions \cite{Be1971,ko1973}. Such is the case of the nematic, 2D-solid and hexatic phases~\cite{strandburgTwodimensionalMelting1988,NaLyPo1992,Nelson1983,Gennes1993,Nelson2002,BaMe2013,Me2015,Me2017,Thorneywork2017} in a vast class of different physical systems like ultra-thin magnetic films \cite{AbKa1995,MeLuDi2020,Hu2020}, quasi two dimensional copolymer systems \cite{Li2006,Gla_2007}), vortex matter in two dimensional superconductors \cite{Gui2009,DiMe2017,Ro2019} and many others \cite{Me2015,Me2017}.  

The study of these kind of phases has its origin in a series of pioneering works by Berezinskii, Kosterlitz, Thouless {(BKT)}, Halperin, Nelson and Young \cite{Be1970,Be1971,ko1973,Ne1978,NeHa1979,OsHa1981,Yo1979,ToNe1981}. In these works it was established the central role of the thermal induced process of pair unbinding and proliferation of topological defects in the disruption of the quasi long-range order.
{From the technical point of view the development of this research area is historically tied with the development of a proper Renormalization Group (RG) theory capable to describe the proliferation of the topological excitations at high-temperature. The success of this RG scheme granted the 2016 Nobel prize in physics\,\cite{kosterlitz2016kosterlitz}.}

Despite this huge success, important questions remain open today {in the physics of 2D transitions.}
{Most of these questions originate from the difficulty to properly estimate non-universal properties, such as the transition temperature, in the wide range of diverse physical system, where topological defects unbinding occurs. Indeed, while the BKT RG scheme yields a phenomenological description of the coarse grained topological variables, the connection between their properties and the actual microscopic theory at hand is often unfeasible. Thus, a natural question arises ``How can we maintain a close connection between the microscopic details of the system and its BKT description?''. In the context of the melting of two-dimensional crystals, one can also ask ``How  can we construct a RG scheme which accounts for} the variety of scenarios observed in simulations and experiments that include first order phase transitions \cite{KaKr2015,BeKr2011,LiCi2020,LiCiPi2020,Chui1983}?". {The answers to these questions are the topic of the present work.}

In order to access non-universal properties, first, it is necessary to build RG equations maintaining full connection with the microscopic model. Second, good estimates of the relevant energy of the defects are needed to capture the temperature scale of the corresponding phase transition~\cite{BeCa2007,MaDe2020}. And lastly, the effects of smooth fluctuations on the ground state should be taken into account, since close to the melting temperature they could produce a significant deviation of the effective microscopic rigidity from its bare or zero temperature value~\cite{ko1973,Bruun2016,MeLuDi2020}. {Following this route it has been possible to estimate the properties of the topological phase transition in diverse XY model configurations\,\cite{defenu2017nonperturbative,MaDe2020,colcelli2020finite,GiDe2021,Defenu2022}. In few specific cases the Self-Consistent Harmonic Approximation (SCHA) has been used to calculate the effective rigidity as a function of temperature and simultaneously this information used as an input of the RG equations in order to estimate the melting temperature better than RG or variational Mean Field alone\,\cite{GiDe2021,BeCa2007}.} 

In this context, the construction of a similar calculation scheme for the study of the melting process of two dimensional crystals still constitutes an unexplored route. The construction of such method shall provide not only good estimates for the melting temperature of two dimensional crystals but shall also yield a self-consistent approach providing the correct qualitative behavior of the phases below and above the transition. {This last property constitutes a substantial advantage} with respect to the several implementations of Density Functional Theory (DFT), which have been used to tackle the melting problem in two and three dimensional crystals,\cite{Ryzhov1995, Likos2006, Prestipino2012}. {These traditional approaches} in the vast majority of cases consider that the 2D solid phase can be treated as a periodic phase which breaks translational symmetry, yielding a major shortcoming that in general produces a wrong description of the critical properties of the melting transition, or even predict incorrectly the nature of the phase transition itself. {In light of the previous discussion, a theory of 2D melting, that we intend to construct, represents a major advancement towards the understanding of melting in two-dimensional crystals, as it will be capable to explain} the diversity of melting scenarios observed in numerical simulations \cite{LiCi2020,LiCiPi2020,KaKr2015,BeKr2011,Xu2016,Glotzer2017,Xu2021,Khali2021,Toledano2021}.

In this work we implement a Self-Consistent Harmonic technique that retains total connection with the microscopic model. This allows us to determine the effective elastic Lamé's coefficients as a function of the density of particles and the temperature. A first mean field estimate of the melting temperature then can be found as the moment when the effective transversal elastic rigidity goes to zero. Contrasting with all other DFT techniques in two dimensions, this kind of mean field calculation has the distinctive virtue of being able to describe properly the qualitative behavior of the phases below and above the melting transition. To improve the mean field results, we use the obtained values for the effective elastic coefficients as an input into the RG equations for the melting transition of a two-dimensional solid. This produces a strong correction to the phase boundary of the 2D solid phase. To test the quality of this method that incorporates the SCHA into the RG theory we performed extensive overdamped Langevin simulations to determine an accurate melting curve for the two dimensional solid phase. The obtained analytical results shows an impressive agreement with the simulational results.  

\section{Self-consistent Harmonic Approximation for the 2D Solid Melting}
We consider a classical system of particles in two dimensions interacting through a Gaussian pairwise potential of the form $V(r)=V_0\exp(-r^2/r_0^2)$, known as the Gaussian-Core Model (GCM) \cite{langFluidSolidPhases2000, stillingerGaussianCoreModel1981, prestipinoHexaticPhaseTwoDimensional2011}. The parameters $r_0$ and $V_0$ represent the range and the intensity of our soft-core potential, respectively. The partition function of this model can be written directly in terms of the configurational integral, which is obtained after integrating over all momenta of the particles:
\begin{align}
    Z=&\frac{1}{N!}\int \left(\prod_i \frac{d^2\mathbf{r}_i}{\Lambda^2}\right) \mathrm{e}^{-\beta \sum_{i<j} V(|\mathbf{r}_i-\mathbf{r}_j|)},
\end{align}
where $\Lambda\equiv h/\sqrt{2\pi m k_B T}$ is the de Broglie thermal wave length and $N$ the number of particles. Having in mind that our goal is to study the melting of the two-dimensional solid phase it is natural to consider that each particle will effectively explore a region of finite "volume" around its equilibrium position. This consideration allows us to write the position of each particle $\textbf{r}_i$ as $\textbf{R}_i+\textbf{u}_i$, where $\textbf{R}_i$ represent the equilibrium lattice position of the $i$-particle and $\textbf{u}_i$ its relative position respect $\textbf{R}_i$. 

For the pair interaction potential we are considering, the ground-state configuration is given by a triangular lattice with spacing $a=(2/\sqrt{3}\rho)^{1/2}$, where $\rho$ stand for the density of particles in two dimensions. In this way the equilibrium positions of the particles can be represented as $\textbf{R}_j=a\left(n \textbf{e}_1+m \textbf{e}_2\right)$, with $n$ and $m$ integers and the basis vectors of the lattice taken as $\textbf{e}_1=(1,0)$ and $\textbf{e}_2=(-1/2,{\sqrt{3}}/{2})$.

Once we include the combinatorial factor of distributing the $N$ particles of the system on the $N$ sites of the solid, the partition function can be rewritten as:
\begin{align}
    Z= & \int \prod_i \frac{d^2\mathbf{u}_i}{\Lambda^2} \mathrm{e}^{- \frac{\beta}{2}\sum_{i\neq j} V(|\mathbf{R}_i-\mathbf{R}_j+\mathbf{u}_i-\mathbf{u}_j|)}.
\end{align}

This last expression allows us to define our effective Hamiltonian as $H\equiv \sfrac{1}{2} \sum_{i\neq j} V(|\mathbf{R}_i-\mathbf{R}_j+\mathbf{u}_i-\mathbf{u}_j|)$. Before we proceed to implement the Self-Consistent Harmonic Approximation (SCHA) in order to obtain the melting curve of the 2D solid phase, it is worth mentioning that the usual mean-field approaches devised as a variational theory on the local density profile fails to identify the 2D solid phase. Instead, depending on the pair interaction potential, such theory can predict at most a phase with broken translational symmetry in two dimensions - which is forbidden by the Mermim-Wagner theorem. Moreover, in our case, since the Fourier transform of our pair interaction potential is positive-definite, even such a theory would fail to identify the existence of the solid phase.
Furthermore, the GCM have a density versus temperature reentrant hexagonal solid phase in its low temperature regime \cite{stillingerGaussianCoreModel1981} and the melting of the solid phase occurs through a two-step process with an intermediate hexatic phase existing in a very narrow temperature region \cite{prestipinoHexaticPhaseTwoDimensional2011}.

Now we proceed with the implementation of the SCHA. The Fourier transform of the field $\mathbf{u}(\mathbf{R}_j)$ on the triangular lattice is defined as:
\begin{align}
\mathbf{u}(\textbf{R}_j)&=\frac{\sqrt{3}a^2}{2}\int_{BZ} \frac{d^2q}{\left(2\pi\right)^2} \mathrm{e}^{i\textbf{q}\cdot\textbf{R}_j} \hat{ \mathbf{u}}(\textbf{q}),\\
\hat{\mathbf{u}}(\textbf{q})&=\sum_{j} \mathrm{e}^{-i\textbf{q}\cdot\textbf{R}_j} \mathbf{u}(\textbf{R}_j),
\end{align}
where the momentum integral is performed over the first Brillouin zone ($BZ$) of the triangular lattice. We choose our test Hamiltonian in the usual harmonic form for an hexagonal two dimensional solid (see Appendix \ref{appendix:A}): \begin{align}
\begin{split}
\label{H0}
    H_0= & N\epsilon_0(\rho)+\frac{1}{2}\left(\frac{\sqrt{3}a^2(\rho)}{2}\right) \times \\ & \int_{BZ}\frac{d^2q}{\left(2\pi\right)^2} \left[ f_1(\textbf{q})\left|\hat{ u}_\parallel(\textbf{q})\right|^2 +
    f_2(\textbf{q})\left|\hat{ u}_\perp(\textbf{q})\right|^2\right], 
\end{split}
\end{align}
where $\epsilon_0(\rho)$ represent the ground state energy per particle of the GCM at a given density $\rho$. The dispersion relations $f_1(\mathbf{q})$ and $f_2(\mathbf{q})$ represent the longitudinal and transversal elastic test response functions to be found, in principle, through the minimization of the variational free energy functional. Additionally,  the longitudinal and transversal elastic fields are given as: $\hat{ u}_\parallel(\mathbf{q})=-i\textbf{q}\cdot\hat{\textbf{u}}(\textbf{q})/q$ and $\hat{ u}_\perp(\mathbf{q})=-i\textbf{q}_\perp\cdot\hat{\textbf{u}}(\textbf{q})/q$, with $\textbf{q}_\perp=(-q_y,q_x)$. 

Although the form of the test Hamiltonian may look highly elaborated it is in fact quite intuitive since it coincides with the expansion of $H$ up to second order in powers of $\hat{\textbf{u}}(\textbf{q})$, considering generic dispersion relations for the longitudinal and transversal elastic modes. In the long wave limit ($q\rightarrow 0$), the harmonic theory predicts $f_1(\textbf{q})=(2\mu+\lambda)\textbf{q}^2$ and $f_2(\textbf{q})=\mu \textbf{q}^2$, where $\mu$ and $\lambda$ represent the so-called Lamé's elastic coefficients -- these and other results of harmonic theory for the solid elasticity are reviewed in the Appendix \ref{appendix:A}.   

Now we can proceed with the construction of the variational free energy. As it is well known, the actual free energy of the system is bounded from above by the minimum of the functional:
\begin{equation}
\begin{split}
   F_{var}&= F_0+\left<H\right>_0-\left<H_0\right>_0,
\end{split}
\end{equation}
where $\left<\circ\right>_0=\frac{1}{Z_0}\int\Pi_id\textbf{u}_i\exp(-\beta H_0)\circ$. Given the Gaussian character of our test model it is not hard to conclude that: 
\begin{equation}
\begin{split}
\label{corfunc}
\left<\hat{u}_\parallel(\textbf{q})\hat{u}_\parallel(\textbf{q}')\right>_0&=\frac{(2\pi)^2\delta(\textbf{q}+\textbf{q}')}{\frac{\sqrt{3}a^2}{2}\beta f_1(q)},\\
\left<\hat{u}'_\perp(\textbf{q})\hat{u}'_\perp(\textbf{q}')\right>_0&=\frac{(2\pi)^2\delta(\textbf{q}+\textbf{q}')}{\frac{\sqrt{3}a^2}{2}\beta f_2(q)},
\end{split}
\end{equation}
and consequently: 
\begin{equation}
\begin{split}
\left<H_0\right>_0&=N\epsilon_0+\frac{1}{2}\left(\frac{\sqrt{3}a^2}{2}\right)\times\\
&\int \frac{d^2q}{(2\pi)^2}\left[f_1(\mathbf{q}) \left<|\hat u_\parallel(\mathbf{q})|^2\right>_0+f_2(\mathbf{q}) \left<|\hat u_\perp(\mathbf{q})|^2\right>_0\right]\\
&=N\epsilon_0+N k_B T.
\end{split}
\end{equation}

Now we focus on the determination of the free energy of the test model $F_0$, which can be obtained from the corresponding partition function $Z_0$. In this case the quadratic form of the test Hamiltonian (\ref{H0}) allows for a direct integration of the Gaussian degrees of freedom, obtaining:
\begin{equation}
    Z_0=\mathrm{e}^{-\beta N\epsilon_0}\left(\prod_\mathbf{q}\frac{2\pi}{\beta\Lambda^2\sqrt{f_1(\mathbf{q})f_2(\mathbf{q})}}\right),
\end{equation}
which lead us to a free energy of the test Hamiltonian of the form:
\begin{equation}
\begin{split}
F_0&=-\frac{1}{\beta}\log{Z_0}=N\epsilon_0+\frac{1}{\beta}\sum_\textbf{q}\log\left(\frac{\beta\Lambda^2}{2\pi}\sqrt{f_1(\mathbf{q}) f_2(\mathbf{q})}\right)\\
&=N\epsilon_0+\frac{N}{\beta\rho}\int_{BZ}\frac{d^2q}{(2\pi)^2} \log\left[\frac{\beta\Lambda^2}{2\pi}\sqrt{f_1(\mathbf{q}) f_2(\mathbf{q})}\right].
\end{split}
\end{equation}

To proceed with the construction of our variational free energy, we should now determine $\left<H\right>_0$. Considering the translational symmetry of the solid phase, it is possible to write:
\begin{equation}
    \left<H\right>_0=\frac{N}{2}\sum_{\mathbf{R}_i\neq \mathbf{0}} \left< V\left(\mathbf{R}_i+\mathbf{u}(\mathbf{R}_i) - \mathbf{u}(\mathbf{0})\right)\right>_0,
\end{equation}
where $\left\{\mathbf{R}_i\right\}$ corresponds to the equilibrium positions of the particles in a hexagonal solid at a fixed density. 
At the same time, the average interaction energy between two particles linked to lattice sites separated by a vector $\mathbf{R}$ is given by:
\begin{align}
\label{intpot}
\left<V\left(\mathbf{R}+\mathbf{u}(\mathbf{R}) - \mathbf{u}(\mathbf{0})\right)\right>_0 & =\int\frac{d^2 q}{(2\pi)^2} \mathrm{e}^{i\mathbf{q}\cdot\mathbf{R}}
g_{\mathbf{q}}(\mathbf{R}) \hat V(q),
\end{align}
where we have used the standard form of the Fourier transform in two dimensions to decouple the average process of the potential $V(\textbf{r})$.
Here $g_{\mathbf{q}}(\mathbf{R})\equiv \left< \mathrm{e}^{i\mathbf{q}\cdot(\mathbf{u}(\mathbf{R})-\mathbf{u}(0))}\right>_0$, represent the positional correlation function calculated with the Boltzmann measure of $H_0$. Considering now the Gaussian nature of the stochastic variables $\mathbf{u}(\mathbf{R})$, we can express $g_{\mathbf{q}}(\mathbf{R})$ as:
\begin{widetext}
\begin{equation}
    \begin{split}\label{gaussianmeans1}
    g_{\textbf{q}}(\textbf{R}) &= \exp\left[-\frac{1}{2}\sum_{\alpha,\beta} q_\alpha q_\beta \left<(u_\alpha(\textbf{R})-u_\alpha(0))(u_\beta(\textbf{R})-u_\beta(0))\right>_0\right]\\
    &= \exp\left[-\frac{1}{2}\sum_{\alpha,\beta} q_\alpha q_\beta \left(\frac{\sqrt{3}a^2}{2}\right)^2\int\frac{d^2k d^2k'}{(2\pi)^2}\left(\mathrm{e}^{i\textbf{k}\cdot\textbf{R}}-1\right)\left(\mathrm{e}^{i\textbf{k}'\cdot\textbf{R}}-1\right)\left<u_\alpha(\textbf{k})u_\beta(\textbf{k}')\right>_0 \right]\\
    &= \exp\left\{-\frac{1}{2}\frac{\sqrt{3}a^2}{2}\int_{BZ} \frac{d^2 k}{(2\pi)^2}
    2 \left(1-\cos(\textbf{k}\cdot\textbf{R})\right)\left[\frac{(\textbf{k}\cdot \textbf{q})^2}{k^2}\left<|\hat{u}_\parallel(\textbf{k})|^2\right>_0+\frac{(\textbf{k}_\perp\cdot \textbf{q})^2}{k^2}\left<|\hat{u}_\perp(\textbf{k})|^2\right>_0\right] \right\}\\
    &=\exp\left\{-\frac{1}{\beta\rho}\int_{BZ}\frac{d^2k}{(2\pi)^2}\left(1-\cos(\textbf{k}\cdot\textbf{R})\right)\left[\frac{(\textbf{k}\cdot \textbf{q})^2}{k^2}\frac{1}{f_1(\mathbf{k})}+\frac{(\textbf{k}_\perp\cdot \textbf{q})^2}{k^2}\frac{1}{f_2(\mathbf{k})}\right]\right\}.
    \end{split}
\end{equation}
\end{widetext}
At this point we can verify that in the limit of zero temperature ($\beta\rightarrow\infty$) the positional correlation function converges to one, as expected for the perfectly ordered crystal.

The expression in Eq.~(\ref{gaussianmeans1}), once inserted in Eq.~(\ref{intpot}), complete the construction of the variational free energy functional in terms of $f_1(\mathbf{q})$ and $f_2(\mathbf{q})$.      
In this way, the variational free energy per particle in units of $k_BT$ can be written as:
\begin{equation}
\begin{split}
\beta f_{var}[f_1(\textbf{q}),&f_2(\textbf{q})]=\frac{1}{\rho}\int_{BZ}\frac{d^2q}{(2\pi)^2}\log\left[\frac{\beta\Lambda^2}{2\pi}\sqrt{f_1(\textbf{q}) f_2(\textbf{q})}\right]\\
&-1+\frac{\beta}{2}\sum_{\mathbf{R}_j\neq\mathbf{0}} \int \frac{d^2 q}{(2\pi)^2} \mathrm{e}^{i\mathbf{q}\cdot\mathbf{R}_j}g_{\mathbf{q}}(\mathbf{R}_j) \hat V(q).
\end{split}
\end{equation}

Demanding $\delta f_{var}/\delta f_1(\textbf{q}_0)=0$  and $\delta f_{var}/\delta f_2(\textbf{q}_0)=0$, we obtain the following set of integral equations for $f_1(\textbf{q})$ and $f_2(\textbf{q})$:
\begin{equation}
\begin{split}
\label{eqf}
f_1(\textbf{q}_0)=\sum_{
\textbf{R}_j\neq0}&\left(\cos(\textbf{q}_0\cdot \textbf{R}_j)-1\right) \times \\
&\int \frac{d^2q}{(2\pi)^2} \mathrm{e}^{i\textbf{q}\cdot\textbf{R}_{j}}\hat V(q)
g_{\textbf{q}}(\textbf{R}_j)  \frac{(\textbf{q}_0\cdot \textbf{q})^2}{q_0^2}\\
f_2(\textbf{q}_0)=\sum_{
\textbf{R}_j\neq0}&\left(\cos(\textbf{q}_0\cdot \textbf{R}_j)-1\right)\times \\
& \int \frac{d^2q}{(2\pi)^2} \mathrm{e}^{i\textbf{q}\cdot\textbf{R}_{j}}\hat V(q)
g_{\textbf{q}}(\textbf{R}_j)  \frac{(\textbf{q}_0^{\perp}\cdot \textbf{q})^2}{q_0^2},
\end{split}
\end{equation}
where $\textbf{q}_0=(q_{0,x},q_{0,y})$, $\textbf{q}_0^{\perp}=(-q_{0,y},q_{0,x})$ and 
$g_{\textbf{q}}(\textbf{R}_j)$ is defined in Eq. (\ref{gaussianmeans1}). The numerical solution of this set of integral equations is a quite difficult task. However, it is expected that the long distance elastic properties of the system is captured by low momentum behavior of $f_1(\textbf{q})$ and $f_2(\textbf{q})$.
This can be used to follow a simpler approach to determine the boundary in the phase diagram of the 2D solid phase. 
Considering the symmetries of the solid phase under study, we already know that in the low momentum regime ($\textbf{q}\rightarrow 0$) the leading order term of the functions $f_1(\textbf{q})$ and $f_2(\textbf{q})$ is proportional to $q^2$. This feature can be used to derive a system of equations for the effective Lamé coefficients of the solid, which we define as:
\begin{equation}
\begin{split}
r_1(T) &= \frac{1}{2}\,\lim_{q\rightarrow0}(\partial^2 f_1(q)/\partial q^2) \equiv 2\mu(T)+\lambda(T)\\
r_2(T) &=\frac{1}{2}\,\lim_{q\rightarrow0}(\partial^2 f_2(q)/\partial q^2)\equiv \mu(T).    
\end{split}
\end{equation}
In this way, considering the form of Eq.~(\ref{eqf}), we can conclude that the elastic coefficients satisfy the following set of equations: 
\begin{equation}
\begin{split}\label{syst} 
r_1&=-\dfrac{1}{2}\sum_{\textbf{R}_i\neq0} (\textbf{e}\cdot\textbf{R}_i)^2 \int\dfrac{d^2q}{(2\pi)^2} \mathrm{e}^{i\textbf{q}\cdot\textbf{R}_i}\hat V(q) g_{\mathbf{q}}(\mathbf{R}_i) (\textbf{e}\cdot \textbf{q})^2,\\
r_2&=-\dfrac{1}{2}\sum_{\textbf{R}_i\neq0} (\textbf{e}\cdot\textbf{R}_i)^2 \int\dfrac{d^2q}{(2\pi)^2} \mathrm{e}^{i\textbf{q}\cdot\textbf{R}_i}\hat V(q) g_{\mathbf{q}}(\mathbf{R}_i) (\textbf{e}_\perp\cdot \textbf{q})^2,   
\end{split}
\end{equation}
where $\textbf{e}$ can be taken as $(1,0)$ and $\textbf{e}_\perp$ as $(0,1)$. Once we have obtained an exact system of equations for $r_1$ and $r_2$ it is natural to approximate $f_1$ and $f_2$ by its low momentum form ($f_1(\mathbf{k})=r_1k^2$ and $f_2(\mathbf{k})=r_2k^2$) in the calculus of $g_{\mathbf{q}}(\mathbf{R})$. This consideration lead us, after a simple but lengthy calculation, to:
\begin{equation}\label{eiquw1w2}
    g_{\mathbf{q}}(\mathbf{R})=\exp\left[-\omega_\parallel(\textbf{R}) q_\parallel^2-\omega_\perp(\textbf{R}) q_\perp^2\right],
\end{equation}
where $q_\parallel$ and $q_\perp$ represent the components of $\textbf{q}$ parallel and perpendicular to $\textbf{R}$, and the coefficients $\omega_\parallel(\textbf{R})$ and $\omega_\perp(\textbf{R})$ depend on $r_1$ and $r_2$ in the following way:
\begin{equation}
\label{eiquw1w2_2}
\begin{split}
\omega_\parallel(\textbf{R})&=\frac{1}{\beta\rho}\left(\frac{A_\parallel(\textbf{R})}{r_1}+\frac{A_\perp(\textbf{R})}{r_2}\right)\\
\omega_\perp(\textbf{R})&=\frac{1}{\beta\rho}\left(\frac{A_\perp(\textbf{R})}{r_1}+\frac{A_\parallel(\textbf{R})}{r_2}\right).
\end{split}
\end{equation}
and the functions $A_{\parallel}(\textbf{R})$ and $A_{\perp}(\textbf{R})$ are defined as the following integrals over the first Brillouin zone of the hexagonal crystal:
\begin{equation}
\begin{split}
A_\parallel(\textbf{R})&=\int_{BZ}\frac{d^2k}{(2\pi)^2}\frac{1-\cos(\textbf{k}\cdot\textbf{R})}{k^4}\frac{(\textbf{k}\cdot\textbf{R})^2}{R^2}\\
A_\perp(\textbf{R})&=\int_{BZ}\frac{d^2k}{(2\pi)^2}\frac{1-\cos(\textbf{k}\cdot\textbf{R})}{k^4}\frac{(\textbf{k}\cdot\textbf{R}_\perp)^2}{R^2}.
\end{split}
\end{equation}

Here it is important to mention that, in the determination of $g_\textbf{q}(\textbf{R})$, we have neglected the cross term proportional to $q_{\parallel}q_\perp$ in the argument of the exponential function. This approximation is very well justified since, depending on the direction of $\textbf{R}$, the corresponding coefficient is either zero, or a small and rapidly decreasing function of $R$ whose highest value is of the order of $1\%$ of the other coefficients in the quadratic form.

In this way the obtained expression for $g_{\mathbf{q}}(\mathbf{R})$ in Eqs.~(\ref{eiquw1w2}) and (\ref{eiquw1w2_2}) allows us to close the system of Eqs.~(\ref{syst}) for $r_1(T)$ and $r_2(T)$. Such system can now be used to determine the effective Lamé coefficients as they are affected by phonon fluctuations. We can then identify the 2D solid phase as the region in the phase diagram in which the effective shear modulus $r_2(T)>0$. 

\subsection{Application to the Gaussian-Core Model (GCM)}

As mentioned earlier in this work we focus on the analytical study of the GCM for which $\hat V(q)=\pi V_0\mathrm{e}^{-q^2r_0^2/4}$. 
The Gaussian form of this potential allows the analytical integration of the right-hand side of equations (\ref{syst}), leading us to following system of equations for $r_1(T)$ and $r_2(T)$:
\begin{equation}
\begin{split}\label{consistencyrelationsexact}
&r_{1,2}
=-\sum_\textbf{R} \frac{R_x^2 \mathrm{e}^{-\frac{R^2}{1+4\omega_\parallel}}}{32R^2(\frac{1}{4}+\omega_\parallel)^{5/2}(\frac{1}{4}+\omega_\perp)^{3/2}}\times\\
&\left[{2R_{y,x}^2 (\frac{1}{4}+\omega_\parallel)^2-R_{x,y}^2\left(R^2-2(\frac{1}{4}+\omega_\parallel)\right)(\frac{1}{4}+\omega_\perp)}\right].
\end{split}
\end{equation}
where the sum over $\mathbf{R}$, as in eq.~(\ref{syst}), is performed over the hexagonal lattice with spacing $a(\rho)$, and $\omega_\parallel$ and $\omega_\perp$ represents the functions of $\textbf R$ given by eq.~(\ref{eiquw1w2_2}). Finally, the coefficients $A_{\parallel}(\textbf R)$ and $A_{\perp}(\textbf R)$ 
entering in the system (\ref{consistencyrelationsexact})
are determine numerically for each value of $\mathbf{R}$. Now we can proceed with the numerical solution of the system for $r_1$ and $r_2$ at any density and temperature.

\begin{figure}[ht!]
\adjustbox{trim=0.6cm 0.8cm 0.6cm 0.6cm}{\resizebox{1.0\linewidth}{!}{\begingroup
  \makeatletter
  \providecommand\color[2][]{\GenericError{(gnuplot) \space\space\space\@spaces}{Package color not loaded in conjunction with
      terminal option `colourtext'}{See the gnuplot documentation for explanation.}{Either use 'blacktext' in gnuplot or load the package
      color.sty in LaTeX.}\renewcommand\color[2][]{}}\providecommand\includegraphics[2][]{\GenericError{(gnuplot) \space\space\space\@spaces}{Package graphicx or graphics not loaded}{See the gnuplot documentation for explanation.}{The gnuplot epslatex terminal needs graphicx.sty or graphics.sty.}\renewcommand\includegraphics[2][]{}}\providecommand\rotatebox[2]{#2}\@ifundefined{ifGPcolor}{\newif\ifGPcolor
    \GPcolorfalse
  }{}\@ifundefined{ifGPblacktext}{\newif\ifGPblacktext
    \GPblacktexttrue
  }{}\let\gplgaddtomacro\g@addto@macro
\gdef\gplbacktext{}\gdef\gplfronttext{}\makeatother
  \ifGPblacktext
\def\colorrgb#1{}\def\colorgray#1{}\else
\ifGPcolor
      \def\colorrgb#1{\color[rgb]{#1}}\def\colorgray#1{\color[gray]{#1}}\expandafter\def\csname LTw\endcsname{\color{white}}\expandafter\def\csname LTb\endcsname{\color{black}}\expandafter\def\csname LTa\endcsname{\color{black}}\expandafter\def\csname LT0\endcsname{\color[rgb]{1,0,0}}\expandafter\def\csname LT1\endcsname{\color[rgb]{0,1,0}}\expandafter\def\csname LT2\endcsname{\color[rgb]{0,0,1}}\expandafter\def\csname LT3\endcsname{\color[rgb]{1,0,1}}\expandafter\def\csname LT4\endcsname{\color[rgb]{0,1,1}}\expandafter\def\csname LT5\endcsname{\color[rgb]{1,1,0}}\expandafter\def\csname LT6\endcsname{\color[rgb]{0,0,0}}\expandafter\def\csname LT7\endcsname{\color[rgb]{1,0.3,0}}\expandafter\def\csname LT8\endcsname{\color[rgb]{0.5,0.5,0.5}}\else
\def\colorrgb#1{\color{black}}\def\colorgray#1{\color[gray]{#1}}\expandafter\def\csname LTw\endcsname{\color{white}}\expandafter\def\csname LTb\endcsname{\color{black}}\expandafter\def\csname LTa\endcsname{\color{black}}\expandafter\def\csname LT0\endcsname{\color{black}}\expandafter\def\csname LT1\endcsname{\color{black}}\expandafter\def\csname LT2\endcsname{\color{black}}\expandafter\def\csname LT3\endcsname{\color{black}}\expandafter\def\csname LT4\endcsname{\color{black}}\expandafter\def\csname LT5\endcsname{\color{black}}\expandafter\def\csname LT6\endcsname{\color{black}}\expandafter\def\csname LT7\endcsname{\color{black}}\expandafter\def\csname LT8\endcsname{\color{black}}\fi
  \fi
    \setlength{\unitlength}{0.0500bp}\ifx\gptboxheight\undefined \newlength{\gptboxheight}\newlength{\gptboxwidth}\newsavebox{\gptboxtext}\fi \setlength{\fboxrule}{0.5pt}\setlength{\fboxsep}{1pt}\definecolor{tbcol}{rgb}{1,1,1}\begin{picture}(6376.00,4676.00)\gplgaddtomacro\gplbacktext{\csname LTb\endcsname \put(866,2618){\makebox(0,0)[r]{\strut{}$0$}}\put(866,3148){\makebox(0,0)[r]{\strut{}$0.1$}}\put(866,3677){\makebox(0,0)[r]{\strut{}$0.2$}}\put(866,4207){\makebox(0,0)[r]{\strut{}$0.3$}}\put(956,2468){\makebox(0,0){\strut{}}}\put(1647,2468){\makebox(0,0){\strut{}}}\put(2337,2468){\makebox(0,0){\strut{}}}\put(3028,2468){\makebox(0,0){\strut{}}}\put(2807,3995){\makebox(0,0)[l]{\strut{}$\rho r_0^2$}}\put(1020,4067){\makebox(0,0)[l]{\strut{}$(a)$}}\put(1020,2384){\makebox(0,0)[l]{\strut{}$(c)$}}\put(3634,4067){\makebox(0,0)[l]{\strut{}$(b)$}}\put(3634,2384){\makebox(0,0)[l]{\strut{}$(d)$}}}\gplgaddtomacro\gplfronttext{\csname LTb\endcsname \put(453,3412){\rotatebox{-270}{\makebox(0,0){\strut{}$\mu/V_0$}}}\put(2199,2393){\makebox(0,0){\strut{}$ $}}\csname LTb\endcsname \put(2978,3791){\makebox(0,0)[r]{\strut{}$0.18$}}\csname LTb\endcsname \put(2978,3596){\makebox(0,0)[r]{\strut{}$0.26$}}\csname LTb\endcsname \put(2978,3401){\makebox(0,0)[r]{\strut{}$0.34$}}\csname LTb\endcsname \put(2978,3206){\makebox(0,0)[r]{\strut{}$0.42$}}\csname LTb\endcsname \put(2978,3011){\makebox(0,0)[r]{\strut{}$0.50$}}}\gplgaddtomacro\gplbacktext{\csname LTb\endcsname \put(3480,2618){\makebox(0,0)[r]{\strut{}}}\put(3480,3148){\makebox(0,0)[r]{\strut{}}}\put(3480,3677){\makebox(0,0)[r]{\strut{}}}\put(3480,4207){\makebox(0,0)[r]{\strut{}}}\put(4009,2468){\makebox(0,0){\strut{}}}\put(4594,2468){\makebox(0,0){\strut{}}}\put(5179,2468){\makebox(0,0){\strut{}}}\put(5764,2468){\makebox(0,0){\strut{}}}}\gplgaddtomacro\gplfronttext{\csname LTb\endcsname \put(3337,3412){\rotatebox{-270}{\makebox(0,0){\strut{}$ $}}}\put(4813,2393){\makebox(0,0){\strut{}$ $}}}\gplgaddtomacro\gplbacktext{\csname LTb\endcsname \put(866,935){\makebox(0,0)[r]{\strut{}$0$}}\put(866,1376){\makebox(0,0)[r]{\strut{}$0.75$}}\put(866,1818){\makebox(0,0)[r]{\strut{}$1.5$}}\put(866,2259){\makebox(0,0)[r]{\strut{}$2.25$}}\put(956,785){\makebox(0,0){\strut{}$0$}}\put(1647,785){\makebox(0,0){\strut{}$0.025$}}\put(2337,785){\makebox(0,0){\strut{}$0.05$}}\put(3028,785){\makebox(0,0){\strut{}$0.075$}}}\gplgaddtomacro\gplfronttext{\csname LTb\endcsname \put(363,1729){\rotatebox{-270}{\makebox(0,0){\strut{}$(2\mu+\lambda)/V_0$}}}\put(2199,560){\makebox(0,0){\strut{}$k_B T/V_0$}}}\gplgaddtomacro\gplbacktext{\csname LTb\endcsname \put(3480,935){\makebox(0,0)[r]{\strut{}}}\put(3480,1376){\makebox(0,0)[r]{\strut{}}}\put(3480,1818){\makebox(0,0)[r]{\strut{}}}\put(3480,2259){\makebox(0,0)[r]{\strut{}}}\put(4009,785){\makebox(0,0){\strut{}$0.2$}}\put(4594,785){\makebox(0,0){\strut{}$0.4$}}\put(5179,785){\makebox(0,0){\strut{}$0.6$}}\put(5764,785){\makebox(0,0){\strut{}$0.8$}}}\gplgaddtomacro\gplfronttext{\csname LTb\endcsname \put(3337,1729){\rotatebox{-270}{\makebox(0,0){\strut{}$ $}}}\put(4813,560){\makebox(0,0){\strut{}$\rho r_0^2$}}\csname LTb\endcsname \put(5485,1296){\makebox(0,0)[r]{\strut{}$T=0$}}\csname LTb\endcsname \put(5485,1101){\makebox(0,0)[r]{\strut{}$T=T_m$}}}\gplbacktext
    \put(0,0){\includegraphics[width={318.80bp},height={233.80bp}]{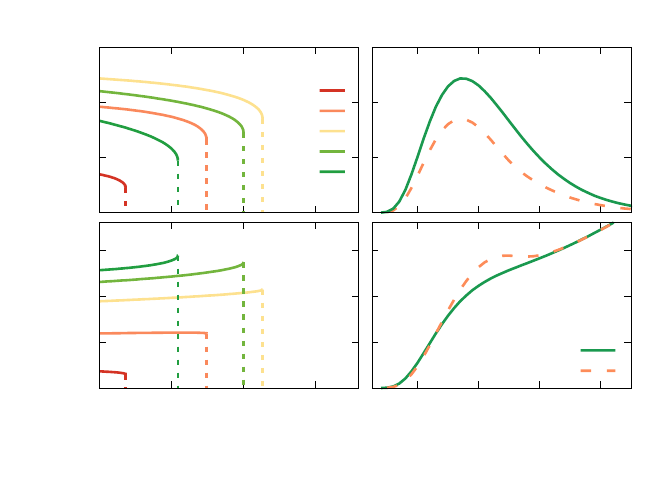}}\gplfronttext
  \end{picture}\endgroup
 }}
    \caption{Behaviour of the Lamé coeficients $\mu$ and $2\mu+\lambda$, resulting from the variational approach as function of the temperature, for some specific density values. The temperature ranges between zero and the critical value $T_m$, at which the $\mu(T)$ function has a infinite slope. Comparison of the Lamé coefficients at $T=0$ and $T=T_m$. The solid curves represent the ground state values and the dashed lines represent the value at $T=T_m$.}
    \label{fig1}
\end{figure}

In the first column of Fig.~\ref{fig1} we show the numerical solution for $\mu$ and $2\mu+\lambda$ as a function of temperature at several fixed densities of particles. As we can see, the transversal elastic coefficient, or shear modulus, $\mu(T)$ displays the typical behavior of the order parameter in a BKT-like transition with an abrupt decay when temperature approaches  the melting temperature ($T_m$). In terms of densities, we are able to observe the re-entrant behavior of $\mu(T)$ already expected from previous numerical simulations results \cite{prestipinoHexaticPhaseTwoDimensional2011}. In the other hand, the longitudinal elastic coefficient $2\mu(T)+\lambda(T)$ grows steadily with the density, a expected behaviour for any stable solid. Additionally, we observe that this quantity is an increasing function of temperature in the region of densities where $\mu$ takes its maximum value, as can be seen in Fig.~\ref{fig1}-b. This unusual behavior is related with the fact that in this region of densities the fluctuations in the position of particles tends to increase the effective repulsion between them.

In the second column of Fig.~(\ref{fig1}), we show a comparison between the values of $\mu$ and $2\mu + \lambda$, at zero and at the meting temperature, as the density is increased. We observe that the value of $\mu$ decreases with temperatures at all densities, while $(2\mu+\lambda)$ displays a weak non-monotonical behavior with temperature. 

Finally, the melting curve in the temperature versus density plane can be observed in Fig.~(\ref{figfinal}) in green. A detailed analysis of this phase diagram for the GCM and its comparison with equivalent results using different techniques will be presented in the following sections. However, we can anticipate that direct comparison with simulation results lead us to conclude that this mean-field technique overestimates significantly the maximum melting temperature of the model. Nevertheless, the SCHA developed in this work has the merit of being a novel mean field calculation describing properly the ground-state properties of the system as well as the qualitative behavior of the 2D solid phase and its melting.

\section{Defect mediated phase transition and RG relations}

In order to improve the agreement between the mean field results and the computational results 
is necessary to take into account the effects of the relevant topological defects, known as dislocations.
The proliferation of dislocations is quite effective in disrupting the periodic order of the solid phase as the melting process occurs increasing temperature. The theory for describing the defect mediated melting transition in two dimensions, also known as KTHNY theory, was provided in a series of foundational works by Toner, Halperin, Nelson and Young \cite{NeHa1979,HaNe1978,ToNe1981,Yo1979} in which RG equations are obtained for the renormalized Young's modulus and fugacity of the dislocations, respectively.  

The RG system of equation for the Young's modulus in units of $k_BT$ ($K(l)$) and the defect's fugacity ($y(l)$) is given by:
\begin{align}
\nonumber
\label{RGsys}
\frac{dK^{-1}}{dl}&=\frac{3}{2}\pi y^2(l) \mathrm{e}^{K(l)/8\pi}\left[I_0\left(\frac{K(l)}{8\pi}\right)-\frac{1}{2}I_1\left(\frac{K(l)}{8\pi}\right)\right],\\
\frac{dy}{dl}&=\left[2-\frac{K(l)}{8\pi}\right]y(l)+2\pi y^2(l)\mathrm{e}^{K(l)/16\pi}I_0\left(\frac{K(l)}{8\pi}\right),
\end{align}
with the initial conditions given by the bare values of $K$ and $y$, i.e. $K(0)=\frac{4\mu(\mu+\lambda)}{(2\mu+\lambda)k_BT}$ and $y(0)=\exp(-E_c/k_BT)$. Here $\mu$ and $\lambda$ represent the Lamé's coefficients and $E_c$ represent the energy of an isolated defect. Within RG theory the melting temperature corresponds to the lowest temperature at which $K(l\rightarrow+\infty)=0$, signaling that the long distance effective rigidity of the system goes to zero. 

To use the RG equations for detecting the melting transition the energy of the relevant defects in the melting process should be provided. In our case we estimate such quantity using the harmonic elastic theory to calculate half of the energy corresponding to a pair of conjugate dislocations. The obtained result depends on the orientation of the Burguer's vector of the dislocations ($\textbf{e}$) with respect to the underlying lattice, and weakly on the orientation of the distance vector between the dislocations of the pair $\textbf{d}$. Since there is a continuum of possible configurations for the dislocation pair, the energy of the dislocation is estimated as the average energy between all configurations maintaining the minimum possible length of the dislocation pair. In this way the energy of a dislocation in units of $k_BT$ is estimated to be:
\begin{align}
\label{Ec}
    \frac{E_c}{k_BT}&=K(0)\int_{BZ} \frac{d^2q}{(2\pi)^2}\frac{\left<(\textbf{q}_\perp\cdot\textbf{e})^2\right>\left< \sin(\textbf{q}\cdot\textbf{d}/2)^2\right>}{q^4} \nonumber\\
    &\approx0.072K(0).
\end{align}

An important consideration to reach this result is related to the minimum distance between a pair of stable dislocations. Numerical simulations performed for the GCM model lead us to the conclusion that such distance is higher than the lattice spacing. One useful way to obtain a pair of stable dislocations in this system is to subtract a particle from a perfect crystal configuration and leave the system to relax to a new stable configuration exhibiting a pair of dislocations separated by a distance that can be roughly estimated to $d=\sqrt{3}a$ \cite{Fisher1979}. More details on how to reach this conclusion and calculate the energy of a dislocation pair can be found in Appendix~\ref{appendix:B}.

We realize now that, since the initial condition of the RG flow equations are completely determined by the value of the bare Young's modulus in units of thermal energy $K(0)$, the melting transition will occur at a certain specific value of the parameter $K(0)$. In this case the direct numerical solution of the system of Eqs.~(\ref{RGsys}) allow us to conclude that the melting transition occurs approximately at $K(0)_c \approx 23.922\pi$. It is important to mention that within the KTHNY theory, it is well established that the effective Young's modulus at the critical point $K(l\rightarrow\infty)$ is equal to $16 \pi$. However the corresponding value of the bare Young's modulus $K(0)$ it is not an universal quantity and its value at the critical point depends on the specific energy cost of the relevant defects. The value of $K(0)_c$ can now be used to build the melting curve considering the calculated Lamé's coefficients, $\mu(T)$ and $\lambda(T)$, dressed by phonon fluctuations. The melting temperature ($T_m$) of the 2D solid phase at a given density is determined from the self-consistency relation $4\mu(\rho,T_m)(\mu(\rho,T_m)+\lambda(\rho,T_m))/\left((2\mu(\rho,T_m)+\lambda(\rho,T_m))k_BT_m\right)=23.922\pi$. 

\begin{figure}[ht]
    \centering
    \adjustbox{trim=0.4cm 0.3cm 0.0cm 0.0cm}{\resizebox{1.0\linewidth}{!}{\begingroup
  \makeatletter
  \providecommand\color[2][]{\GenericError{(gnuplot) \space\space\space\@spaces}{Package color not loaded in conjunction with
      terminal option `colourtext'}{See the gnuplot documentation for explanation.}{Either use 'blacktext' in gnuplot or load the package
      color.sty in LaTeX.}\renewcommand\color[2][]{}}\providecommand\includegraphics[2][]{\GenericError{(gnuplot) \space\space\space\@spaces}{Package graphicx or graphics not loaded}{See the gnuplot documentation for explanation.}{The gnuplot epslatex terminal needs graphicx.sty or graphics.sty.}\renewcommand\includegraphics[2][]{}}\providecommand\rotatebox[2]{#2}\@ifundefined{ifGPcolor}{\newif\ifGPcolor
    \GPcolorfalse
  }{}\@ifundefined{ifGPblacktext}{\newif\ifGPblacktext
    \GPblacktexttrue
  }{}\let\gplgaddtomacro\g@addto@macro
\gdef\gplbacktext{}\gdef\gplfronttext{}\makeatother
  \ifGPblacktext
\def\colorrgb#1{}\def\colorgray#1{}\else
\ifGPcolor
      \def\colorrgb#1{\color[rgb]{#1}}\def\colorgray#1{\color[gray]{#1}}\expandafter\def\csname LTw\endcsname{\color{white}}\expandafter\def\csname LTb\endcsname{\color{black}}\expandafter\def\csname LTa\endcsname{\color{black}}\expandafter\def\csname LT0\endcsname{\color[rgb]{1,0,0}}\expandafter\def\csname LT1\endcsname{\color[rgb]{0,1,0}}\expandafter\def\csname LT2\endcsname{\color[rgb]{0,0,1}}\expandafter\def\csname LT3\endcsname{\color[rgb]{1,0,1}}\expandafter\def\csname LT4\endcsname{\color[rgb]{0,1,1}}\expandafter\def\csname LT5\endcsname{\color[rgb]{1,1,0}}\expandafter\def\csname LT6\endcsname{\color[rgb]{0,0,0}}\expandafter\def\csname LT7\endcsname{\color[rgb]{1,0.3,0}}\expandafter\def\csname LT8\endcsname{\color[rgb]{0.5,0.5,0.5}}\else
\def\colorrgb#1{\color{black}}\def\colorgray#1{\color[gray]{#1}}\expandafter\def\csname LTw\endcsname{\color{white}}\expandafter\def\csname LTb\endcsname{\color{black}}\expandafter\def\csname LTa\endcsname{\color{black}}\expandafter\def\csname LT0\endcsname{\color{black}}\expandafter\def\csname LT1\endcsname{\color{black}}\expandafter\def\csname LT2\endcsname{\color{black}}\expandafter\def\csname LT3\endcsname{\color{black}}\expandafter\def\csname LT4\endcsname{\color{black}}\expandafter\def\csname LT5\endcsname{\color{black}}\expandafter\def\csname LT6\endcsname{\color{black}}\expandafter\def\csname LT7\endcsname{\color{black}}\expandafter\def\csname LT8\endcsname{\color{black}}\fi
  \fi
    \setlength{\unitlength}{0.0500bp}\ifx\gptboxheight\undefined \newlength{\gptboxheight}\newlength{\gptboxwidth}\newsavebox{\gptboxtext}\fi \setlength{\fboxrule}{0.5pt}\setlength{\fboxsep}{1pt}\begin{picture}(5102.00,3400.00)\gplgaddtomacro\gplbacktext{\csname LTb\endcsname \put(980,640){\makebox(0,0)[r]{\strut{}$0$}}\put(980,1280){\makebox(0,0)[r]{\strut{}$0.015$}}\put(980,1920){\makebox(0,0)[r]{\strut{}$0.03$}}\put(980,2559){\makebox(0,0)[r]{\strut{}$0.045$}}\put(980,3199){\makebox(0,0)[r]{\strut{}$0.06$}}\put(1100,440){\makebox(0,0){\strut{}$0.1$}}\put(1585,440){\makebox(0,0){\strut{}$0.2$}}\put(2071,440){\makebox(0,0){\strut{}$0.3$}}\put(2556,440){\makebox(0,0){\strut{}$0.4$}}\put(3042,440){\makebox(0,0){\strut{}$0.5$}}\put(3527,440){\makebox(0,0){\strut{}$0.6$}}\put(4013,440){\makebox(0,0){\strut{}$0.7$}}\put(4498,440){\makebox(0,0){\strut{}$0.8$}}}\gplgaddtomacro\gplfronttext{\csname LTb\endcsname \put(334,1919){\rotatebox{-270}{\makebox(0,0){\strut{}$k_B T/V_0$}}}\put(2920,140){\makebox(0,0){\strut{}$\rho r_0^2$}}\csname LTb\endcsname \put(4052,2856){\makebox(0,0)[r]{\strut{}SCHA+RG}}\csname LTb\endcsname \put(4052,2596){\makebox(0,0)[r]{\strut{}SCHA}}\csname LTb\endcsname \put(4052,2336){\makebox(0,0)[r]{\strut{}MD}}}\gplbacktext
    \put(0,0){\includegraphics{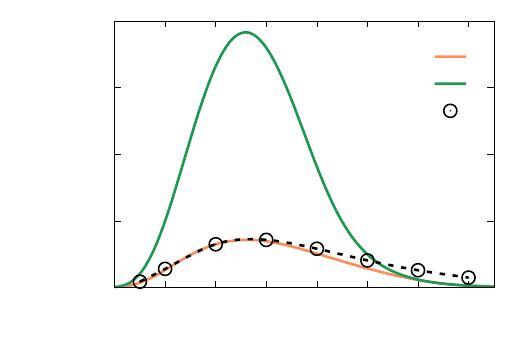}}\gplfronttext
  \end{picture}\endgroup
 }}
\caption{Phase diagram for simulations are represented by black circles and dashed line (cubic splines used as guide to the eyes). The result for variational mean field is presented by a green line and renormalization group + variational approach in orange.}
    \label{figfinal}
\end{figure}

The result of this procedure is shown in Fig.~\ref{figfinal}, which can be compared with our Molecular Dynamics (MD) simulations results for the 2D solid melting temperature, discussed further in the next section. The agreement between the analytical and computational results is quite impressive, indicating that the proposed method not only captures well the phenomenology of the described phase transition but also produce a precise estimation of the melting transition.

\section{Numerical simulations}

In order to confront the theoretical phase diagrams found in the previous sections, we have performed molecular dynamics (MD) simulations of the GCM in a NVT or canonical ensemble. It is worth mentioning that the analytical results of this work are derived in the same ensemble, while previous simulation results for this model are available in the NPT ensemble \cite{prestipinoHexaticPhaseTwoDimensional2011}. We are interested in sampling the equilibrium configurations, so we have employed the Langevin equation in the overdamped limit to simulate the behavior of the system in contact with a heat bath:
\begin{align}
\gamma \dot{\textbf{r}}_i=-\sum_j \nabla_i V(&\left|\textbf{r}_i-\textbf{r}_j\right|)+\sqrt{2k_B T\gamma}\,\xi_i(t),
\end{align}
where $T$ is the temperature of the thermal bath and $\gamma$ is the viscosity. We have chosen to measure the timescales in units of $\gamma$ and the temperature scales in units of $V_0/k_B$. The random force $\xi_i(t)$ is a white noise with $\left<\xi_i(t)\right>=0$ and $\left<\xi_i(t)\xi_j(t')\right>=\delta_{ij}\delta(t-t')$.

The simulations have been performed using the Heun algorithm to integrate numerically the $N$ stochastic differential equations of motion \cite{gardIntroductionStochasticDifferential1988}, for which we used a time step $dt/\gamma=0.1$. 
In order to accelerate the relaxation to thermal equilibrium at all temperatures, we have used the parallel tempering technique \cite{Hukushima1996,sugitaReplicaexchangeMolecularDynamics1999},
considering sets of two types of initial conditions, one corresponding to a liquid (disordered) and another to a crystalline hexagonal lattice. This allows to define a criterion to identify thermal equilibrium at each temperature as the stage at which all sets of configurations attain the same stationary state. We have used a number of particles ranging from 1024 up to 8100 particles.

\begin{figure}[ht!]
    \centering
    \adjustbox{trim=0.0cm 0.0cm 0.5cm 0.0cm}{\resizebox{0.9\linewidth}{!}{\begingroup
\definecolor{myred}{HTML}{b50b7a}\definecolor{myblue}{HTML}{017324}
  \makeatletter
  \providecommand\color[2][]{\GenericError{(gnuplot) \space\space\space\@spaces}{Package color not loaded in conjunction with
      terminal option `colourtext'}{See the gnuplot documentation for explanation.}{Either use 'blacktext' in gnuplot or load the package
      color.sty in LaTeX.}\renewcommand\color[2][]{}}\providecommand\includegraphics[2][]{\GenericError{(gnuplot) \space\space\space\@spaces}{Package graphicx or graphics not loaded}{See the gnuplot documentation for explanation.}{The gnuplot epslatex terminal needs graphicx.sty or graphics.sty.}\renewcommand\includegraphics[2][]{}}\providecommand\rotatebox[2]{#2}\@ifundefined{ifGPcolor}{\newif\ifGPcolor
    \GPcolorfalse
  }{}\@ifundefined{ifGPblacktext}{\newif\ifGPblacktext
    \GPblacktexttrue
  }{}\let\gplgaddtomacro\g@addto@macro
\gdef\gplbacktext{}\gdef\gplfronttext{}\makeatother
  \ifGPblacktext
\def\colorrgb#1{}\def\colorgray#1{}\else
\ifGPcolor
      \def\colorrgb#1{\color[rgb]{#1}}\def\colorgray#1{\color[gray]{#1}}\expandafter\def\csname LTw\endcsname{\color{white}}\expandafter\def\csname LTb\endcsname{\color{black}}\expandafter\def\csname LTa\endcsname{\color{black}}\expandafter\def\csname LT0\endcsname{\color[rgb]{1,0,0}}\expandafter\def\csname LT1\endcsname{\color[rgb]{0,1,0}}\expandafter\def\csname LT2\endcsname{\color[rgb]{0,0,1}}\expandafter\def\csname LT3\endcsname{\color[rgb]{1,0,1}}\expandafter\def\csname LT4\endcsname{\color[rgb]{0,1,1}}\expandafter\def\csname LT5\endcsname{\color[rgb]{1,1,0}}\expandafter\def\csname LT6\endcsname{\color[rgb]{0,0,0}}\expandafter\def\csname LT7\endcsname{\color[rgb]{1,0.3,0}}\expandafter\def\csname LT8\endcsname{\color[rgb]{0.5,0.5,0.5}}\else
\def\colorrgb#1{\color{black}}\def\colorgray#1{\color[gray]{#1}}\expandafter\def\csname LTw\endcsname{\color{white}}\expandafter\def\csname LTb\endcsname{\color{black}}\expandafter\def\csname LTa\endcsname{\color{black}}\expandafter\def\csname LT0\endcsname{\color{black}}\expandafter\def\csname LT1\endcsname{\color{black}}\expandafter\def\csname LT2\endcsname{\color{black}}\expandafter\def\csname LT3\endcsname{\color{black}}\expandafter\def\csname LT4\endcsname{\color{black}}\expandafter\def\csname LT5\endcsname{\color{black}}\expandafter\def\csname LT6\endcsname{\color{black}}\expandafter\def\csname LT7\endcsname{\color{black}}\expandafter\def\csname LT8\endcsname{\color{black}}\fi
  \fi
    \setlength{\unitlength}{0.0500bp}\ifx\gptboxheight\undefined \newlength{\gptboxheight}\newlength{\gptboxwidth}\newsavebox{\gptboxtext}\fi \setlength{\fboxrule}{0.5pt}\setlength{\fboxsep}{1pt}\begin{picture}(6376.00,2550.00)\gplgaddtomacro\gplbacktext{\csname LTb\endcsname \put(637,575){\makebox(0,0)[r]{\strut{}$0$}}\put(637,1180){\makebox(0,0)[r]{\strut{}$5000$}}\put(637,1784){\makebox(0,0)[r]{\strut{}$10000$}}\put(637,2389){\makebox(0,0)[r]{\strut{}$15000$}}\put(1101,352){\makebox(0,0){\strut{}$0.0104$}}\put(1836,352){\makebox(0,0){\strut{}$0.0108$}}\put(2571,352){\makebox(0,0){\strut{}$0.0112$}}\put(1101,2549){\makebox(0,0){\strut{}}}\put(1836,2549){\makebox(0,0){\strut{}}}\put(2571,2549){\makebox(0,0){\strut{}}}\put(2204,2171){\makebox(0,0)[l]{\strut{}$N$}}}\gplgaddtomacro\gplfronttext{\csname LTb\endcsname \put(101,1482){\rotatebox{-270}{\makebox(0,0){\strut{}$\textcolor{myred}{\chi_T V_0} \quad \textcolor{myblue}{\chi_6 V_0}$}}}\put(1928,112){\makebox(0,0){\strut{}$k_B T/V_0$}}\csname LTb\endcsname \put(2421,2007){\makebox(0,0)[r]{\strut{}$1024$}}\csname LTb\endcsname \put(2421,1847){\makebox(0,0)[r]{\strut{}$2025$}}\csname LTb\endcsname \put(2421,1687){\makebox(0,0)[r]{\strut{}$4096$}}\csname LTb\endcsname \put(2421,1527){\makebox(0,0)[r]{\strut{}$8100$}}\csname LTb\endcsname \put(2724,2007){\makebox(0,0)[r]{\strut{} }}\csname LTb\endcsname \put(2724,1847){\makebox(0,0)[r]{\strut{} }}\csname LTb\endcsname \put(2724,1687){\makebox(0,0)[r]{\strut{} }}\csname LTb\endcsname \put(2724,1527){\makebox(0,0)[r]{\strut{} }}}\gplgaddtomacro\gplbacktext{\csname LTb\endcsname \put(3155,575){\makebox(0,0)[r]{\strut{}}}\put(3155,1180){\makebox(0,0)[r]{\strut{}}}\put(3155,1784){\makebox(0,0)[r]{\strut{}}}\put(3155,2389){\makebox(0,0)[r]{\strut{}}}\put(3435,352){\makebox(0,0){\strut{}$0.0036$}}\put(4170,352){\makebox(0,0){\strut{}$0.004$}}\put(4906,352){\makebox(0,0){\strut{}$0.0044$}}\put(5641,352){\makebox(0,0){\strut{}$0.0048$}}\put(5737,575){\makebox(0,0)[l]{\strut{}$0$}}\put(5737,1180){\makebox(0,0)[l]{\strut{}$7000$}}\put(5737,1784){\makebox(0,0)[l]{\strut{}$14000$}}\put(5737,2389){\makebox(0,0)[l]{\strut{}$21000$}}\put(3435,2549){\makebox(0,0){\strut{}}}\put(4170,2549){\makebox(0,0){\strut{}}}\put(4906,2549){\makebox(0,0){\strut{}}}\put(5641,2549){\makebox(0,0){\strut{}}}\put(17039,1715){\makebox(0,0)[l]{\strut{}$N$}}\put(824,2216){\makebox(0,0)[l]{\strut{}$(a)$}}\put(3343,2216){\makebox(0,0)[l]{\strut{}$(b)$}}}\gplgaddtomacro\gplfronttext{\csname LTb\endcsname \put(3179,1482){\rotatebox{-270}{\makebox(0,0){\strut{}}}}\put(6297,1482){\rotatebox{-270}{\makebox(0,0){\strut{}$\textcolor{myred}{\chi_T V_0} \quad \textcolor{myblue}{\chi_6 V_0}$}}}\put(4446,112){\makebox(0,0){\strut{}$k_B T/V_0$}}}\gplbacktext
    \put(0,0){\includegraphics{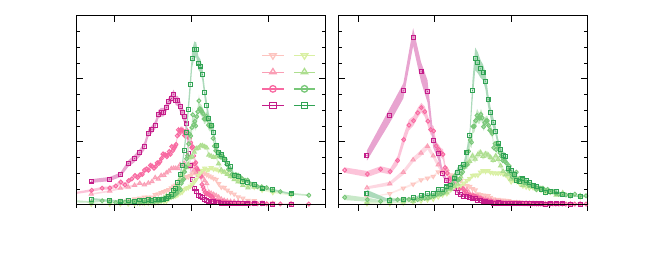}}\gplfronttext
  \end{picture}\endgroup
 }}
\adjustbox{trim=1.5cm 0.3cm 1.cm 0.0cm}{\resizebox{1.05\linewidth}{!}{\begingroup
  \makeatletter
  \providecommand\color[2][]{\GenericError{(gnuplot) \space\space\space\@spaces}{Package color not loaded in conjunction with
      terminal option `colourtext'}{See the gnuplot documentation for explanation.}{Either use 'blacktext' in gnuplot or load the package
      color.sty in LaTeX.}\renewcommand\color[2][]{}}\providecommand\includegraphics[2][]{\GenericError{(gnuplot) \space\space\space\@spaces}{Package graphicx or graphics not loaded}{See the gnuplot documentation for explanation.}{The gnuplot epslatex terminal needs graphicx.sty or graphics.sty.}\renewcommand\includegraphics[2][]{}}\providecommand\rotatebox[2]{#2}\@ifundefined{ifGPcolor}{\newif\ifGPcolor
    \GPcolorfalse
  }{}\@ifundefined{ifGPblacktext}{\newif\ifGPblacktext
    \GPblacktexttrue
  }{}\let\gplgaddtomacro\g@addto@macro
\gdef\gplbacktext{}\gdef\gplfronttext{}\makeatother
  \ifGPblacktext
\def\colorrgb#1{}\def\colorgray#1{}\else
\ifGPcolor
      \def\colorrgb#1{\color[rgb]{#1}}\def\colorgray#1{\color[gray]{#1}}\expandafter\def\csname LTw\endcsname{\color{white}}\expandafter\def\csname LTb\endcsname{\color{black}}\expandafter\def\csname LTa\endcsname{\color{black}}\expandafter\def\csname LT0\endcsname{\color[rgb]{1,0,0}}\expandafter\def\csname LT1\endcsname{\color[rgb]{0,1,0}}\expandafter\def\csname LT2\endcsname{\color[rgb]{0,0,1}}\expandafter\def\csname LT3\endcsname{\color[rgb]{1,0,1}}\expandafter\def\csname LT4\endcsname{\color[rgb]{0,1,1}}\expandafter\def\csname LT5\endcsname{\color[rgb]{1,1,0}}\expandafter\def\csname LT6\endcsname{\color[rgb]{0,0,0}}\expandafter\def\csname LT7\endcsname{\color[rgb]{1,0.3,0}}\expandafter\def\csname LT8\endcsname{\color[rgb]{0.5,0.5,0.5}}\else
\def\colorrgb#1{\color{black}}\def\colorgray#1{\color[gray]{#1}}\expandafter\def\csname LTw\endcsname{\color{white}}\expandafter\def\csname LTb\endcsname{\color{black}}\expandafter\def\csname LTa\endcsname{\color{black}}\expandafter\def\csname LT0\endcsname{\color{black}}\expandafter\def\csname LT1\endcsname{\color{black}}\expandafter\def\csname LT2\endcsname{\color{black}}\expandafter\def\csname LT3\endcsname{\color{black}}\expandafter\def\csname LT4\endcsname{\color{black}}\expandafter\def\csname LT5\endcsname{\color{black}}\expandafter\def\csname LT6\endcsname{\color{black}}\expandafter\def\csname LT7\endcsname{\color{black}}\expandafter\def\csname LT8\endcsname{\color{black}}\fi
  \fi
    \setlength{\unitlength}{0.0500bp}\ifx\gptboxheight\undefined \newlength{\gptboxheight}\newlength{\gptboxwidth}\newsavebox{\gptboxtext}\fi \setlength{\fboxrule}{0.5pt}\setlength{\fboxsep}{1pt}\begin{picture}(5384.00,3400.00)\gplgaddtomacro\gplbacktext{\csname LTb\endcsname \put(980,640){\makebox(0,0)[r]{\strut{}$0$}}\put(980,1493){\makebox(0,0)[r]{\strut{}$0.004$}}\put(980,2346){\makebox(0,0)[r]{\strut{}$0.008$}}\put(980,3199){\makebox(0,0)[r]{\strut{}$0.012$}}\put(1100,440){\makebox(0,0){\strut{}$0.1$}}\put(1623,440){\makebox(0,0){\strut{}$0.2$}}\put(2146,440){\makebox(0,0){\strut{}$0.3$}}\put(2669,440){\makebox(0,0){\strut{}$0.4$}}\put(3192,440){\makebox(0,0){\strut{}$0.5$}}\put(3715,440){\makebox(0,0){\strut{}$0.6$}}\put(4238,440){\makebox(0,0){\strut{}$0.7$}}\put(4761,440){\makebox(0,0){\strut{}$0.8$}}\put(1257,2986){\makebox(0,0)[l]{\strut{}$(c)$}}\put(2251,1920){\makebox(0,0)[l]{\strut{}2D Solid}}\put(3297,2773){\makebox(0,0)[l]{\strut{}Hexatic}}\put(1241,2346){\makebox(0,0)[l]{\strut{}Fluid}}\put(4207,2212){\makebox(0,0)[l]{\strut{}Fluid}}}\gplgaddtomacro\gplfronttext{\csname LTb\endcsname \put(310,1919){\rotatebox{-270}{\makebox(0,0){\strut{}$k_B T/V_0$}}}\put(3061,140){\makebox(0,0){\strut{}$\rho r_0^2$}}}\gplbacktext
    \put(0,0){\includegraphics{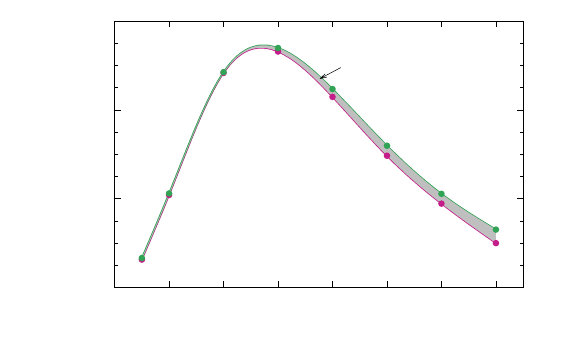}}\gplfronttext
  \end{picture}\endgroup
 }}
    \caption{Susceptibilities of the order parameters for increasing number of particles, for {\it (a)} $\rho r_0^2 = 0.4$ and {\it (b)} $\rho r_0^2 = 0.7$. The susceptibility associated to the translational order parameter ($\chi_T$) in presented in magenta and the corresponding to the orientational order parameter ($\chi_6$) in green. The width of the filled curves represent the statistical uncertainty of the data. {\it (c)}
    Phase diagram of the GCM from MD simulations in NVT ensemble. The points in green are the estimates for the hexatic-liquid phase boundary and points in magenta are the analog for the 2D solid-hexatic phase boundary. The lines are guides to the eyes. The narrow region in grey corresponds to the hexatic phase.}
    \label{fig:suscept}
\end{figure}

In order to characterize the phase diagram of the system, we measure two relevant and well established order parameters associated to the translational ($\psi_T$) and bond-orientational orders ($\psi_6$) \cite{Nelson1983}. They are defined as:
\begin{align}
  \Psi_T&=\frac{1}{N}\left|\sum_j^N \mathrm{e}^{i \textbf k_0\cdot\textbf r}\right|,\\
  \Psi_6&=\frac{1}{N}\left|\sum_j^N \frac{1}{N_{nn}(j)}\sum_l^{N_{nn}(j)} \mathrm{e}^{6 i \theta_{jl}}\right|,
\end{align}
where $\textbf k_0$ represents the characteristic wave vector of the reciprocal lattice, estimated from the position of the peaks in the structure factor. Additionally, $N_{nn}(j)$ stand for the number of nearest neighbours of the particle $j$ and $\theta_{jl}$ is the bond angle between particles $j$ and $l$, both determined from a Delaunay triangulation.

The goal of our computational study is to address the construction of the density versus temperature phase diagram for the GCM. In particular, we focus in the estimation of the 2D solid melting temperature. Furthermore, it is well established in previous works that the solid-liquid melting in the GCM occurs through an intermediate hexatic phase, present in a quite slim region next to the 2D solid phase boundary. Also, in this work we estimate the phase boundaries of the hexatic phase.

The signatures of the transitions are more pronounced if we study the corresponding order parameter susceptibilities, defined as:
\begin{equation}
    \chi_\alpha\equiv\frac{1}{k_B}\frac{\partial \Psi_\alpha}{\partial T}=\frac{N}{k_B T}\left[\left<\Psi_\alpha^2\right>-\left<\Psi_\alpha\right>^2\right].
\end{equation}

In order to estimate the 2D solid melting temperature we observed the maximum of the translational order parameter susceptibility ($\chi_T$), whereas for the hexatic melting temperature, the maximum of orientational order parameter susceptibility ($\chi_6$). These quantities are shown as a function of temperature in the first row of Fig.~\ref{fig:suscept} for an increasing number of particles and for two values of the density. The corresponding melting temperatures can be estimated by extrapolating linearly the location of the maxima as a function of ${N}^{-1/2}$. It is important to mention that considering 
the temperature corresponding to the maximum of the susceptibilities for the largest system size would produce a visually  indistinguishable phase diagram in comparison with the one presented above. The resulting phase diagram is shown in Fig.~\ref{fig:suscept}.c, where we can notice that the extension of the hexatic phase is quite narrow in temperature and increases progressively with density from $\rho r_0^2>0.4$. For the comparison between simulational and analytical results in Fig.~\ref{figfinal} we have considered the magenta curve from Fig.~\ref{fig:suscept}.c, which corresponds to the boundary of the 2D solid phase.

\subsection{Melting criteria from Young modulus measurements}

One of the conclusions of our analytical results in the previous section is that the bare Young modulus $K(0)$ in units of $k_BT$ at the melting temperature takes approximately the value of $23.9\pi$ for the GCM. This value is a consequence of the microscopic model constructed to estimate the core energy of the dislocations, shown in detail in  Appendix \ref{appendix:B}. Conversely, the $K(l\to \infty)=16\pi$ criterion for the renormalized Young modulus is an universal prediction from the KTHNY theory valid for any model in principle. In this scenario, a microscopic calculation of the bare and renormalized Young modulus would be an important and independent validation of our analytical predictions.

The Young modulus can be measured in numerical simulation by well established techniques, such as the one that measures the elastic properties from unperturbed equilibrium distances between particles, 
shown in early works from Squire {\it et al}$\,$ \cite{squireIsothermalElasticConstants1969}. This method uses a quadratic expansion in the deformation field for the Helmholtz free energy density $f$, in order to extract the elastic tensor components $C_{ijkl}=\partial^2 f/\partial \eta_{ij}\partial \eta_{kl}|_{\eta=0}$, where $\eta_{ij}$ represent a small strain tensor perturbation in the equilibrium positions of the particles. These elastic constants can be expressed as a sum of different contributions $C_{xyxy} = C_{44} = C^{(B)} + \rho k_B T - C_{44}^{(F)}$ and $C_{xxyy} = C_{12} = C^{(B)}- C_{12}^{(F)}$. Here, the Born term is the same for both of these constants and can be written as:
\begin{align}
    C^{(B)} = \frac{1}{V}\left<\sum_{i<j}\left(V^{\prime\prime}(r_{ij})-\frac{V^{\prime}(r_{ij})}{r_{ij}}\right)\frac{x^2_{ij}y^2_{ij}}{r_{ij}^2}\right>, 
\end{align}
while the fluctuation terms are given by:
\begin{align}
      C^{(F)}_{44} &= \frac{V}{k_B T} \left[ \left<\sigma_{xy}^2\right>-\left<\sigma_{xy} \right>^2 \right] \label{cf44}\\
      C^{(F)}_{12} &= \frac{V}{k_B T} \left[\left<\sigma_{xx}\sigma_{yy}\right>-\left<\sigma_{xx}\right>\left<\sigma_{yy}\right> \right] \label{cf12},
\end{align}
where the stress tensor is defined as:
\begin{align}
    \sigma_{\alpha\beta} = \frac{1}{V}\sum_{i<j}V^{\prime}(r_{ij})\frac{r^{\alpha}_{ij}r^{\beta}_{ij}}{r_{ij}},
    \label{stress}
\end{align}
where $\alpha, \beta$ are the Cartesian components of the distance $\textbf{r}_{ij}$ between pairs of particles. In order to estimate the Young's modulus, we define the shear modulus as $G = C_{44} -P$ and the bulk modulus as $B = C_{44}+C_{12}$, where the pressure $P$ can be calculated using the virial expression. In this way, the Young's modulus can be found as:
   \begin{align}
     K = \frac{4a^2}{k_BT}\frac{GB}{G+B}
     \label{young-sims}
   \end{align}
where $a=(2/\sqrt{3}\rho)^{1/2}$ is the triangular lattice spacing. This quantity has been measured in two-dimensional melting simulation studies using the same technique depicted above \cite{Abraham1981, Tobochnik1982, Allen1983}, as well as using other methods \cite{Broughton1982, Toxvaerd1983, Morales1994}. 

The fluctuation contributions to the elastic constants given by Eqs. (\ref{cf44}) and (\ref{cf12}) are variances of the intensive variables $\sigma_{\alpha\beta}$. These variances are expected to behave as $1/N$, in order to produce finite values of $C_{44}^{(F)}$ and $C_{12}^{(F)}$ in the thermodynamic limit.
Within the SCHA the variables $r^\alpha_{ij}$ in Eq. (\ref{stress}) have a Gaussian probability distribution. In general, such equation can be rewritten in terms of the Fourier transform of the pair potential $V(r)$, which allow us to translate the nontrivial $r_{ij}$-dependence of the kernel of $\sigma_{\alpha\beta}$ to an expression that contains $r_{ij}$ only in the argument of complex exponentials. The stochastic average of these kind of quantities decay exponentially with the quadratic fluctuation of $r_{ij}$. In this way, the fluctuation contributions to the elastic constants given by Eqs. (\ref{cf44}) and (\ref{cf12}) can be seen as 'variance of variances' that, within the SCHA, are expect to be particularly small.

Within this hypothesis we can identify the Eq.~(\ref{young-sims}) calculated without the contributions from Eqs. (\ref{cf44}) and (\ref{cf12}) as $K(0)$, the bare value of the Young's modulus in units of thermal energy as calculated within the SCHA. Consequently, we can think of this quantity measured in simulations as the effective Young's modulus affected only by phonon fluctuations. From this estimation of $K(0)$ obtained within numerical simulations we can independently verify the melting criterion for the GCM obtained from analytical calculation ($K(0)_c\simeq 23.9\pi$). At the same time, taking into account all contributions to $C_{44}$ and $C_{12}$ in Eq.~(\ref{young-sims}) lead us to the macroscopic effective value of the Young's modulus $K\equiv K(l\to \infty)$ within our simulations. This provides a more standard path for the determination of the melting temperature of the 2D solid phase in numerical simulations, resulting from the criterion $K(T_c)=16\pi$.

\begin{figure}[ht]
    \centering
    \resizebox{0.85\linewidth}{!}{\begingroup
  \makeatletter
  \providecommand\color[2][]{\GenericError{(gnuplot) \space\space\space\@spaces}{Package color not loaded in conjunction with
      terminal option `colourtext'}{See the gnuplot documentation for explanation.}{Either use 'blacktext' in gnuplot or load the package
      color.sty in LaTeX.}\renewcommand\color[2][]{}}\providecommand\includegraphics[2][]{\GenericError{(gnuplot) \space\space\space\@spaces}{Package graphicx or graphics not loaded}{See the gnuplot documentation for explanation.}{The gnuplot epslatex terminal needs graphicx.sty or graphics.sty.}\renewcommand\includegraphics[2][]{}}\providecommand\rotatebox[2]{#2}\@ifundefined{ifGPcolor}{\newif\ifGPcolor
    \GPcolorfalse
  }{}\@ifundefined{ifGPblacktext}{\newif\ifGPblacktext
    \GPblacktexttrue
  }{}\let\gplgaddtomacro\g@addto@macro
\gdef\gplbacktext{}\gdef\gplfronttext{}\makeatother
  \ifGPblacktext
\def\colorrgb#1{}\def\colorgray#1{}\else
\ifGPcolor
      \def\colorrgb#1{\color[rgb]{#1}}\def\colorgray#1{\color[gray]{#1}}\expandafter\def\csname LTw\endcsname{\color{white}}\expandafter\def\csname LTb\endcsname{\color{black}}\expandafter\def\csname LTa\endcsname{\color{black}}\expandafter\def\csname LT0\endcsname{\color[rgb]{1,0,0}}\expandafter\def\csname LT1\endcsname{\color[rgb]{0,1,0}}\expandafter\def\csname LT2\endcsname{\color[rgb]{0,0,1}}\expandafter\def\csname LT3\endcsname{\color[rgb]{1,0,1}}\expandafter\def\csname LT4\endcsname{\color[rgb]{0,1,1}}\expandafter\def\csname LT5\endcsname{\color[rgb]{1,1,0}}\expandafter\def\csname LT6\endcsname{\color[rgb]{0,0,0}}\expandafter\def\csname LT7\endcsname{\color[rgb]{1,0.3,0}}\expandafter\def\csname LT8\endcsname{\color[rgb]{0.5,0.5,0.5}}\else
\def\colorrgb#1{\color{black}}\def\colorgray#1{\color[gray]{#1}}\expandafter\def\csname LTw\endcsname{\color{white}}\expandafter\def\csname LTb\endcsname{\color{black}}\expandafter\def\csname LTa\endcsname{\color{black}}\expandafter\def\csname LT0\endcsname{\color{black}}\expandafter\def\csname LT1\endcsname{\color{black}}\expandafter\def\csname LT2\endcsname{\color{black}}\expandafter\def\csname LT3\endcsname{\color{black}}\expandafter\def\csname LT4\endcsname{\color{black}}\expandafter\def\csname LT5\endcsname{\color{black}}\expandafter\def\csname LT6\endcsname{\color{black}}\expandafter\def\csname LT7\endcsname{\color{black}}\expandafter\def\csname LT8\endcsname{\color{black}}\fi
  \fi
    \setlength{\unitlength}{0.0500bp}\ifx\gptboxheight\undefined \newlength{\gptboxheight}\newlength{\gptboxwidth}\newsavebox{\gptboxtext}\fi \setlength{\fboxrule}{0.5pt}\setlength{\fboxsep}{1pt}\begin{picture}(5102.00,4534.00)\gplgaddtomacro\gplbacktext{\csname LTb\endcsname \put(592,2694){\makebox(0,0)[r]{\strut{}$60$}}\put(592,3097){\makebox(0,0)[r]{\strut{}$70$}}\put(592,3500){\makebox(0,0)[r]{\strut{}$80$}}\put(592,3903){\makebox(0,0)[r]{\strut{}$90$}}\put(592,4306){\makebox(0,0)[r]{\strut{}$100$}}\put(963,2333){\makebox(0,0){\strut{}}}\put(1513,2333){\makebox(0,0){\strut{}}}\put(2063,2333){\makebox(0,0){\strut{}}}\put(2613,2333){\makebox(0,0){\strut{}}}\put(3163,2333){\makebox(0,0){\strut{}}}\put(3713,2333){\makebox(0,0){\strut{}}}\put(4263,2333){\makebox(0,0){\strut{}}}\put(4813,2333){\makebox(0,0){\strut{}}}\put(1,4185){\makebox(0,0)[l]{\strut{}(a)}}\put(1,2292){\makebox(0,0)[l]{\strut{}(b)}}\put(3851,4185){\makebox(0,0)[l]{\strut{}$\rho r_0^2$}}\put(4291,3222){\makebox(0,0)[l]{\strut{}$23.9\pi$}}}\gplgaddtomacro\gplfronttext{\csname LTb\endcsname \put(152,3399){\rotatebox{-270}{\makebox(0,0){\strut{}$K(0)$}}}\csname LTb\endcsname \put(3631,3984){\makebox(0,0)[r]{\strut{}0.2}}\csname LTb\endcsname \put(3631,3824){\makebox(0,0)[r]{\strut{}0.3}}\csname LTb\endcsname \put(3631,3664){\makebox(0,0)[r]{\strut{}0.4}}\csname LTb\endcsname \put(4366,3984){\makebox(0,0)[r]{\strut{}0.5}}\csname LTb\endcsname \put(4366,3824){\makebox(0,0)[r]{\strut{}0.6}}\csname LTb\endcsname \put(4366,3664){\makebox(0,0)[r]{\strut{}0.7}}}\gplgaddtomacro\gplbacktext{\csname LTb\endcsname \put(592,713){\makebox(0,0)[r]{\strut{}$0$}}\put(592,1051){\makebox(0,0)[r]{\strut{}$20$}}\put(592,1389){\makebox(0,0)[r]{\strut{}$40$}}\put(592,1726){\makebox(0,0)[r]{\strut{}$60$}}\put(592,2064){\makebox(0,0)[r]{\strut{}$80$}}\put(592,2402){\makebox(0,0)[r]{\strut{}$100$}}\put(963,384){\makebox(0,0){\strut{}$0.002$}}\put(1513,384){\makebox(0,0){\strut{}$0.004$}}\put(2063,384){\makebox(0,0){\strut{}$0.006$}}\put(2613,384){\makebox(0,0){\strut{}$0.008$}}\put(3163,384){\makebox(0,0){\strut{}$0.01$}}\put(3713,384){\makebox(0,0){\strut{}$0.012$}}\put(4263,384){\makebox(0,0){\strut{}$0.014$}}\put(4813,384){\makebox(0,0){\strut{}$0.016$}}\put(4401,1477){\makebox(0,0)[l]{\strut{}$16\pi$}}}\gplgaddtomacro\gplfronttext{\csname LTb\endcsname \put(152,1473){\rotatebox{-270}{\makebox(0,0){\strut{}$K$}}}\put(2750,144){\makebox(0,0){\strut{}$k_B T/V_0$}}}\gplbacktext
    \put(0,0){\includegraphics{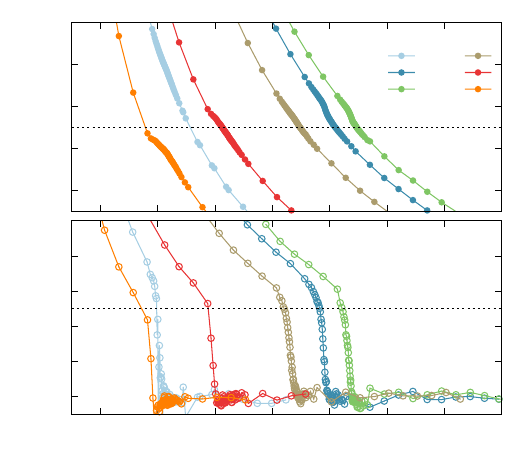}}\gplfronttext
  \end{picture}\endgroup
 }\\
    \adjustbox{trim=0.3cm 0.3cm -0.7cm 0.0cm}{\resizebox{0.90\linewidth}{!}{\begingroup
  \makeatletter
  \providecommand\color[2][]{\GenericError{(gnuplot) \space\space\space\@spaces}{Package color not loaded in conjunction with
      terminal option `colourtext'}{See the gnuplot documentation for explanation.}{Either use 'blacktext' in gnuplot or load the package
      color.sty in LaTeX.}\renewcommand\color[2][]{}}\providecommand\includegraphics[2][]{\GenericError{(gnuplot) \space\space\space\@spaces}{Package graphicx or graphics not loaded}{See the gnuplot documentation for explanation.}{The gnuplot epslatex terminal needs graphicx.sty or graphics.sty.}\renewcommand\includegraphics[2][]{}}\providecommand\rotatebox[2]{#2}\@ifundefined{ifGPcolor}{\newif\ifGPcolor
    \GPcolorfalse
  }{}\@ifundefined{ifGPblacktext}{\newif\ifGPblacktext
    \GPblacktexttrue
  }{}\let\gplgaddtomacro\g@addto@macro
\gdef\gplbacktext{}\gdef\gplfronttext{}\makeatother
  \ifGPblacktext
\def\colorrgb#1{}\def\colorgray#1{}\else
\ifGPcolor
      \def\colorrgb#1{\color[rgb]{#1}}\def\colorgray#1{\color[gray]{#1}}\expandafter\def\csname LTw\endcsname{\color{white}}\expandafter\def\csname LTb\endcsname{\color{black}}\expandafter\def\csname LTa\endcsname{\color{black}}\expandafter\def\csname LT0\endcsname{\color[rgb]{1,0,0}}\expandafter\def\csname LT1\endcsname{\color[rgb]{0,1,0}}\expandafter\def\csname LT2\endcsname{\color[rgb]{0,0,1}}\expandafter\def\csname LT3\endcsname{\color[rgb]{1,0,1}}\expandafter\def\csname LT4\endcsname{\color[rgb]{0,1,1}}\expandafter\def\csname LT5\endcsname{\color[rgb]{1,1,0}}\expandafter\def\csname LT6\endcsname{\color[rgb]{0,0,0}}\expandafter\def\csname LT7\endcsname{\color[rgb]{1,0.3,0}}\expandafter\def\csname LT8\endcsname{\color[rgb]{0.5,0.5,0.5}}\else
\def\colorrgb#1{\color{black}}\def\colorgray#1{\color[gray]{#1}}\expandafter\def\csname LTw\endcsname{\color{white}}\expandafter\def\csname LTb\endcsname{\color{black}}\expandafter\def\csname LTa\endcsname{\color{black}}\expandafter\def\csname LT0\endcsname{\color{black}}\expandafter\def\csname LT1\endcsname{\color{black}}\expandafter\def\csname LT2\endcsname{\color{black}}\expandafter\def\csname LT3\endcsname{\color{black}}\expandafter\def\csname LT4\endcsname{\color{black}}\expandafter\def\csname LT5\endcsname{\color{black}}\expandafter\def\csname LT6\endcsname{\color{black}}\expandafter\def\csname LT7\endcsname{\color{black}}\expandafter\def\csname LT8\endcsname{\color{black}}\fi
  \fi
    \setlength{\unitlength}{0.0500bp}\ifx\gptboxheight\undefined \newlength{\gptboxheight}\newlength{\gptboxwidth}\newsavebox{\gptboxtext}\fi \setlength{\fboxrule}{0.5pt}\setlength{\fboxsep}{1pt}\begin{picture}(5102.00,3400.00)\gplgaddtomacro\gplbacktext{\csname LTb\endcsname \put(784,512){\makebox(0,0)[r]{\strut{}$0$}}\put(784,1421){\makebox(0,0)[r]{\strut{}$0.004$}}\put(784,2330){\makebox(0,0)[r]{\strut{}$0.008$}}\put(784,3239){\makebox(0,0)[r]{\strut{}$0.012$}}\put(1404,352){\makebox(0,0){\strut{}$0.2$}}\put(2453,352){\makebox(0,0){\strut{}$0.4$}}\put(3502,352){\makebox(0,0){\strut{}$0.6$}}\put(4551,352){\makebox(0,0){\strut{}$0.8$}}\put(93,3125){\makebox(0,0)[l]{\strut{}$(c)$}}\put(4026,3057){\makebox(0,0)[l]{\strut{}$K$}}}\gplgaddtomacro\gplfronttext{\csname LTb\endcsname \put(248,1875){\rotatebox{-270}{\makebox(0,0){\strut{}$k_B T/V_0$}}}\put(2846,112){\makebox(0,0){\strut{}$\rho r_0^2$}}\csname LTb\endcsname \put(4305,2794){\makebox(0,0)[r]{\strut{}$23.9\pi$}}\csname LTb\endcsname \put(4305,2586){\makebox(0,0)[r]{\strut{}$16.0\pi$}}}\gplbacktext
    \put(0,0){\includegraphics{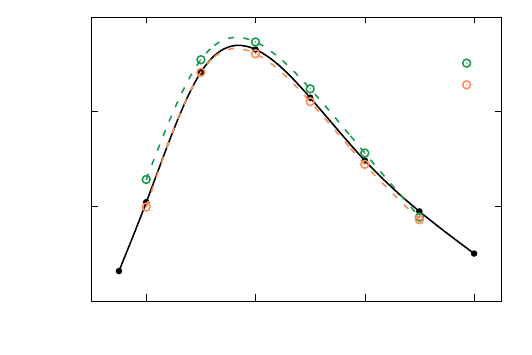}}\gplfronttext
  \end{picture}\endgroup
 }}
\caption{Simulation results for a system with $N$=1024 particles: (a) bare Young's modulus $K(0)$ and (b) Young's modulus $K$ measurements (see main text) for different densities as a function of the temperature. (c) Phase boundary for the 2D solid phase resulting from criteria 
    $K(T_c)=16\pi$ and $K(0)_c\simeq 23.9\pi$. For comparison the full line correspond to the phase boundary as shown in Fig.~\ref{fig:suscept}.(c).}
    \label{figfinalsims}
\end{figure}

Finally, we present in Fig.~\ref{figfinalsims}.(a) the behavior of the bare Young's modulus $K(0)$ together with the reference value of $23.9\pi$ for the melting temperature, and in Fig.~\ref{figfinalsims}.(b) the macroscopic Young's modulus $K$ with its reference value of $16\pi$. By selecting, at each density, the corresponding melting temperature we can draw the estimated 2D solid phase boundary corresponding to each melting criterion. The comparison of these  phase boundaries with the one shown in Fig.\ref{fig:suscept}.(c) is presented in Fig.~\ref{figfinalsims}.(c). As can be observed, the criterion $K(0)_c\simeq 23.9\pi$ for the GCM produces a 2D solid phase boundary that agrees well with the results obtained from more general methods.

\section{Conclusions}

In the present work we have developed a novel technique to study the 2D solid melting transition of the GCM. Although the boundary of this phase is already known from previous computational studies, up to our knowledge, this is the first time that it is reproduced analytically with such level of accuracy. 
In contrast with other MF predictions, we have observed that the SCHA alone is able to predict correctly the qualitative properties of the melting transition, such as the reentrant melting curve and the density for which the melting temperature is higher. In spite of this, only the mechanism of topological defects proliferation related to BKT like transitions is able to explain the significant reduction of the melting temperature when compared with MF predictions. {This mechanism is captured in our approach by performing a two-step calculation where the Young's modulus and the defects fugacity calculated by MF are used as initial condition of the RG flow describing dislocation proliferation and unbinding.}

{Our approach demonstrates how the relative low value of the energy of dislocation pairs causes a drastic reduction of the melting temperature, due to the early proliferation of topological defects. The effect is correctly captured by the RG system in Eq.\,\eqref{RGsys}, once proper values of the defect parameters are introduced}. Our estimate for the energy of the defects can be seen as a leading order approximation, since in general it is expected that higher order contributions will depend on the particle's density and on the Young's modulus itself in a nonlinear way. This is a possible explanation for the small difference observed in Fig.~\ref{figfinal} between the analytical and numerical melting curves at higher densities. A better agreement could be reached by improving the estimation of the energy of the defects, however such involved calculation is beyond the scope of the present work. Additionally, it is important to mention that in the low density regime where the effects of the phonon are weak we verified that, as expected, the melting curves calculated using the zero temperature Young's modulus and using the effective one dressed by phonons produces similar melting curves.

We would like to stress that the presented method is an important step not only in the characterization of the melting transition of two 2D crystals but also to the study of this process in other modulated two dimensional systems, like for magnetic 2D textures\cite{SciRep2020, Barci2013, Mendoza-Coto2012, DiazMendez2011, Nicolao2007}. Finally, it is worth noticing that the developed method has the potential to produce melting scenarios originally not contained in the KTHNY theory. A first order transition melting scenario is in principle a possibility in those cases where the SCHA produces a discontinuous melting transition and the defect energy is high enough for the mean field transition to takes place at lower temperature than the one predicted by RG equations. In this sense, this kind of self-consistent variational plus RG methods can possibly provide a more general framework to explain the variety of melting scenarios observed in different models.

\section*{Acknowledgement}
AM-C acknowledges financial support from Fundação de Amparo à Pesquisa de Santa Catarina, Brazil (Fapesc). AM-C acknowledges hospitality and financial support from MPIPKS. 
ND acknowledges useful discussions with G. Giachetti. This work is supported
by the Deutsche Forschungsgemeinschaft (DFG, German
Research Foundation) under Germany’s Excellence Strategy
EXC2181/1-390900948 (the Heidelberg STRUCTURES
Excellence Cluster).\\

\appendix
\section{Harmonic theory for the solid elasticity}
\label{appendix:A}

In this appendix the elastic properties of the harmonic solid are deduced. The Hamiltonian of a classical system of $N$ particles interacting through a pair potential $V(r)$ is given as:
\begin{equation}
\mathcal{H}=\sum_i \frac{\textbf{p}_i^2}{2 m}+\sum_{i<j} V(|\textbf{r}_i-\textbf{r}_j|),
\end{equation}
and the associated canonical partition function is:
\begin{align}\label{partfunc}
    Z_0&=\frac{1}{N!}\int \left(\prod_i \frac{d^2\textbf{r}_i d^2\textbf{p}_i}{h^2}\right) \mathrm{e}^{-\beta \mathcal{H}} \\
    &=\frac{1}{N!}\int \left(\prod_i \frac{d^2\textbf{r}_i}{\Lambda^2}\right) \mathrm{e}^{-\beta \sum_{i<j} V(|\textbf{r}_i-\textbf{r}_j|)}.
\end{align}

Considering that the ground-state of the system is given by a hexagonal crystal with lattice sites $\textbf{R}_i$, the fluctuations in the particles' positions can be defined as $\textbf{u}_i=\textbf{r}_i-\textbf{R}_i$. We can now expand the potential energy up to second order in $\textbf{u}$, which lead us to:
\begin{align}\label{v0_taylor_expansion}
V_0=&\frac{1}{2}\sum_{i\neq j} V(|\textbf{r}_i-\textbf{r}_j|)=\frac{1}{2}\sum_{i\neq j} V(|\textbf{R}_i-\textbf{R}_j+\textbf{u}_i-\textbf{u}_j|)\nonumber\\
\approx&\frac{1}{2} \sum_{i \neq j} \bigg\{ V(\textbf{R}_i - \textbf{R}_j) + 
\vec\nabla V(\textbf{R}_i - \textbf{R}_j)\cdot \Delta\textbf{u}+\nonumber \\
&+\frac{1}{2}\Delta\textbf{u}^T\cdot\textbf{H}\cdot\Delta\textbf{u}\bigg\}.
\end{align}
Here $\Delta \textbf{u}=\textbf{u}_i-\textbf{u}_j$ and $\textbf{H}$ represent the Hessian matrix of $V(\textbf{R}_i - \textbf{R}_j)$.
The linear term in eq. (\ref{v0_taylor_expansion}) vanishes by symmetry arguments of the ground state. If we use now the definition of Fourier transform on the hexagonal lattice presented in the main text, the interaction energy can be rewritten as:
\begin{widetext}
\begin{equation}
\label{Helastic}
    \begin{split}
    V_0&=N\epsilon_0+\frac{1}{2}\frac{\sqrt{3}a^2}{2}\int_{BZ} \frac{d^2q}{(2\pi)^2}\left(2\left(\frac{1}{2} \partial_x^2 V(R)\right)^{FT}(0)-2\left(\frac{1}{2} \partial_x^2 V(R)\right)^{FT}(\textbf{q})\right)\hat u_x(\textbf{q})\hat u_x(-\textbf{q})\\
    &+\frac{1}{2}\frac{\sqrt{3}a^2}{2}\int_{BZ} \frac{d^2q}{(2\pi)^2}\left(2\left(\frac{1}{2} \partial_y^2 V(R)\right)^{FT}(0)-2\left(\frac{1}{2} \partial_y^2 V(R)\right)^{FT}(\textbf{q})\right)\hat u_y(\textbf{q})\hat u_y(-\textbf{q})\\
    &+\frac{1}{2}\frac{\sqrt{3}a^2}{2}\int_{BZ} \frac{d^2q}{(2\pi)^2}\left(2\left(\partial_{xy} V(R)\right)^{FT}(0)-2\left(\partial_{xy} V(R)\right)^{FT}(\textbf{q})\right)\hat u_x(\textbf{q})\hat u_y(-\textbf{q}),
    \end{split}
\end{equation}
\end{widetext}
where $\epsilon_0$ represent the ground state energy per particle of the system and $(F(\textbf{R}))^{FT}(\textbf{q})\equiv\sum_j \mathrm{e}^{i\textbf{q}\cdot\textbf{R}_j}F(\textbf{R}_j)$. Using this property and expanding the exponential function up the second order in $\textbf{q}$, we arrive at the well known elastic Hamiltonian for a two dimensional hexagonal solid:
\begin{equation}
\label{Helastic1}
    \begin{split}
    V_0=&N\epsilon_0\\
    & +\frac{\sqrt{3}a^2}{4}\int_{BZ} \frac{d^2q}{(2\pi)^2}\left((2\mu+\lambda)q_x^2+\mu q_y^2\right)\hat u_x(\textbf{q})\hat u_x(-\textbf{q})\\
    &+\frac{\sqrt{3}a^2}{4}\int_{BZ} \frac{d^2q}{(2\pi)^2}\left((2\mu+\lambda)q_y^2+\mu q_x^2\right)\hat u_y(\textbf{q})\hat u_y(-\textbf{q})\\
    &+\frac{\sqrt{3}a^2}{4}\int_{BZ} \frac{d^2q}{(2\pi)^2}\left(2(\mu+\lambda)q_xq_y\right)\hat u_x(\textbf{q})\hat u_y(-\textbf{q}).
    \end{split}
\end{equation}
where the Lamé's coefficient ($\mu, \lambda$) can be obtained comparing eqs.~(\ref{Helastic}) and (\ref{Helastic1}). Such relations can be rewritten in real space as:
\begin{equation}
\label{gndstatecoefs}
    \begin{split}
\mu&=\sum_{\mathbf{R}\neq 0} \left(\frac{R_y^2}{2}\right)\left( \frac{R_y^2 V'(R)}{R^3}+\frac{R_x^2 V''(R)}{R^2} \right), \\
2\mu+\lambda&=\sum_{\mathbf{R}\neq 0} \left(\frac{R_x^2}{2}\right)\left( \frac{R_y^2 V'(R)}{R^3}+\frac{R_x^2 V''(R)}{R^2} \right).
    \end{split}
\end{equation}

These are general expressions for the elastic constants at zero temperature for a simple hexagonal crystal, once $V(R)$ is known. 

Finally we would like to rewrite the effective Hamiltonian (\ref{Helastic1}) in its diagonal form, since this is the starting point for the evaluation of any statistical average. We perform an orthogonal transformation introducing the longitudinal $\hat u_\parallel(\textbf{q})$ and perpendicular $\hat u_\perp(\textbf{q})$ components of the deformation field satisfying the following relations:
\begin{equation}
\begin{split}
\hat u_x(\textbf{q}) &= i\frac{q_x}{q}\hat u_\parallel(\textbf{q})-i\frac{q_y}{q}\hat u_\perp(\textbf{q}),\\
\hat u_y(\textbf{q}) &= i\frac{q_y}{q}\hat u_\parallel(\textbf{q})+i\frac{q_x}{q}\hat u_\perp(\textbf{q}).   
\end{split}
\end{equation}
In terms of $\hat u_\parallel(\textbf{q})$ and $\hat u_\perp(\textbf{q})$, the effective Hamiltonian (\ref{Helastic1}) can be recast as:
\begin{equation}
\label{Ham_quadr}
\begin{split}
    V_0 = & N\epsilon_0 +\frac{1}{2}\left(\frac{\sqrt{3}a^2}{2}\right)\int_{BZ}\frac{d^2q}{\left(2\pi\right)^2} \Big[f_1(q)\left|\hat{ u}_\parallel(q)\right|^2\\
    &+f_2(q)\left|\hat{ u}_\perp(q)\right|^2\Big], 
\end{split}
\end{equation}
where $f_1(q)=(2\mu+\lambda)q^2$ and $f_2(q)=\mu q^2$.  

\section{Energy of dislocations and application of the RG theory}
\label{appendix:B}

It is well established in the literature that the long-distance Hamiltonian of a pair of interacting dislocations can be written in Fourier space as \cite{HaNe1978, NeHa1979}:
\begin{align}
\begin{split}
\label{eq:Hd}
    \frac{H_D}{k_BT}&=\frac{-K}{8\pi}\frac{1}{(A_{uc})^2}\int_{BZ} \frac{d^2q}{(2\pi)^2}\sum_{i,j}\hat{\textbf{b}}_i(\textbf{q})\hat{\textbf{b}}_j(-\textbf{q})\\
    &\times\left(-\frac{4\pi}{q^2}\delta_{ij}+4\pi\frac{q_i q_j}{q^4}\right),
\end{split}
\end{align}
where $K=\frac{4\mu(\mu+\lambda)}{(2\mu+\lambda)k_BT}$ is the Young modulus of the crystal in units of $k_BT$, $\hat{\mathbf{b}}(\textbf{q})$ denote the Burger's vector density and $A_{uc}$ represent the area of the unitary cell of the lattice. Although this expression was originally obtained in the continuum limit, to make it well defined a short distance cut-off need to be introduced and in this way the integration over momenta is limited to the first Brilloun zone of the corresponding under-lying lattice of the solid. A dimensional analysis of this equation lead us to the conclusion that $\hat{\textbf{b}}(\textbf{q})$ has the same dimension of  $a^2$, this means that $\textbf{b}(\textbf{r})$ is dimensionless, which is naturally interpreted as  $\textbf{b}(\textbf{r})$ been measured in units of $a$. Here is important to notice that  that due to mathematical convenience all calculations in this appendix are performed in the continuum and consequently the Fourier transform adopt its usual form in two dimensions. 

\begin{figure}[ht!]
    \centering
    \adjustbox{trim=0.5cm 0.5cm 0.6cm 0.cm}{\resizebox{0.45\linewidth}{!}{
        \begingroup
  \makeatletter
  \providecommand\color[2][]{%
    \GenericError{(gnuplot) \space\space\space\@spaces}{%
      Package color not loaded in conjunction with
      terminal option `colourtext'%
    }{See the gnuplot documentation for explanation.%
    }{Either use 'blacktext' in gnuplot or load the package
      color.sty in LaTeX.}%
    \renewcommand\color[2][]{}%
  }%
  \providecommand\includegraphics[2][]{%
    \GenericError{(gnuplot) \space\space\space\@spaces}{%
      Package graphicx or graphics not loaded%
    }{See the gnuplot documentation for explanation.%
    }{The gnuplot epslatex terminal needs graphicx.sty or graphics.sty.}%
    \renewcommand\includegraphics[2][]{}%
  }%
  \providecommand\rotatebox[2]{#2}%
  \@ifundefined{ifGPcolor}{%
    \newif\ifGPcolor
    \GPcolorfalse
  }{}%
  \@ifundefined{ifGPblacktext}{%
    \newif\ifGPblacktext
    \GPblacktexttrue
  }{}%
  \let\gplgaddtomacro\g@addto@macro
  \gdef\gplbacktext{}%
  \gdef\gplfronttext{}%
  \makeatother
  \ifGPblacktext
    \def\colorrgb#1{}%
    \def\colorgray#1{}%
  \else
    \ifGPcolor
      \def\colorrgb#1{\color[rgb]{#1}}%
      \def\colorgray#1{\color[gray]{#1}}%
      \expandafter\def\csname LTw\endcsname{\color{white}}%
      \expandafter\def\csname LTb\endcsname{\color{black}}%
      \expandafter\def\csname LTa\endcsname{\color{black}}%
      \expandafter\def\csname LT0\endcsname{\color[rgb]{1,0,0}}%
      \expandafter\def\csname LT1\endcsname{\color[rgb]{0,1,0}}%
      \expandafter\def\csname LT2\endcsname{\color[rgb]{0,0,1}}%
      \expandafter\def\csname LT3\endcsname{\color[rgb]{1,0,1}}%
      \expandafter\def\csname LT4\endcsname{\color[rgb]{0,1,1}}%
      \expandafter\def\csname LT5\endcsname{\color[rgb]{1,1,0}}%
      \expandafter\def\csname LT6\endcsname{\color[rgb]{0,0,0}}%
      \expandafter\def\csname LT7\endcsname{\color[rgb]{1,0.3,0}}%
      \expandafter\def\csname LT8\endcsname{\color[rgb]{0.5,0.5,0.5}}%
    \else
      \def\colorrgb#1{\color{black}}%
      \def\colorgray#1{\color[gray]{#1}}%
      \expandafter\def\csname LTw\endcsname{\color{white}}%
      \expandafter\def\csname LTb\endcsname{\color{black}}%
      \expandafter\def\csname LTa\endcsname{\color{black}}%
      \expandafter\def\csname LT0\endcsname{\color{black}}%
      \expandafter\def\csname LT1\endcsname{\color{black}}%
      \expandafter\def\csname LT2\endcsname{\color{black}}%
      \expandafter\def\csname LT3\endcsname{\color{black}}%
      \expandafter\def\csname LT4\endcsname{\color{black}}%
      \expandafter\def\csname LT5\endcsname{\color{black}}%
      \expandafter\def\csname LT6\endcsname{\color{black}}%
      \expandafter\def\csname LT7\endcsname{\color{black}}%
      \expandafter\def\csname LT8\endcsname{\color{black}}%
    \fi
  \fi
    \setlength{\unitlength}{0.0500bp}%
    \ifx\gptboxheight\undefined%
      \newlength{\gptboxheight}%
      \newlength{\gptboxwidth}%
      \newsavebox{\gptboxtext}%
    \fi%
    \setlength{\fboxrule}{0.5pt}%
    \setlength{\fboxsep}{1pt}%
\begin{picture}(3400.00,3400.00)%
    \gplgaddtomacro\gplbacktext{%
    }%
    \gplgaddtomacro\gplfronttext{%
    }%
    \gplbacktext
    \put(0,0){\includegraphics{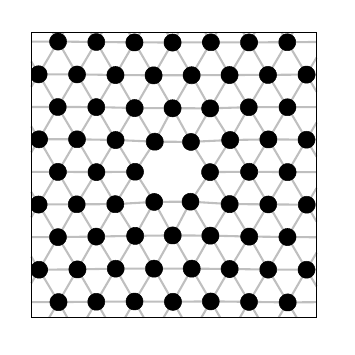}}%
    \gplfronttext
  \end{picture}%
\endgroup
    }}\adjustbox{trim=0.cm 0.5cm 0.5cm 0.5cm}{\resizebox{0.45\linewidth}{!}{\begingroup
  \makeatletter
  \providecommand\color[2][]{\GenericError{(gnuplot) \space\space\space\@spaces}{Package color not loaded in conjunction with
      terminal option `colourtext'}{See the gnuplot documentation for explanation.}{Either use 'blacktext' in gnuplot or load the package
      color.sty in LaTeX.}\renewcommand\color[2][]{}}\providecommand\includegraphics[2][]{\GenericError{(gnuplot) \space\space\space\@spaces}{Package graphicx or graphics not loaded}{See the gnuplot documentation for explanation.}{The gnuplot epslatex terminal needs graphicx.sty or graphics.sty.}\renewcommand\includegraphics[2][]{}}\providecommand\rotatebox[2]{#2}\@ifundefined{ifGPcolor}{\newif\ifGPcolor
    \GPcolorfalse
  }{}\@ifundefined{ifGPblacktext}{\newif\ifGPblacktext
    \GPblacktexttrue
  }{}\let\gplgaddtomacro\g@addto@macro
\gdef\gplbacktext{}\gdef\gplfronttext{}\makeatother
  \ifGPblacktext
\def\colorrgb#1{}\def\colorgray#1{}\else
\ifGPcolor
      \def\colorrgb#1{\color[rgb]{#1}}\def\colorgray#1{\color[gray]{#1}}\expandafter\def\csname LTw\endcsname{\color{white}}\expandafter\def\csname LTb\endcsname{\color{black}}\expandafter\def\csname LTa\endcsname{\color{black}}\expandafter\def\csname LT0\endcsname{\color[rgb]{1,0,0}}\expandafter\def\csname LT1\endcsname{\color[rgb]{0,1,0}}\expandafter\def\csname LT2\endcsname{\color[rgb]{0,0,1}}\expandafter\def\csname LT3\endcsname{\color[rgb]{1,0,1}}\expandafter\def\csname LT4\endcsname{\color[rgb]{0,1,1}}\expandafter\def\csname LT5\endcsname{\color[rgb]{1,1,0}}\expandafter\def\csname LT6\endcsname{\color[rgb]{0,0,0}}\expandafter\def\csname LT7\endcsname{\color[rgb]{1,0.3,0}}\expandafter\def\csname LT8\endcsname{\color[rgb]{0.5,0.5,0.5}}\else
\def\colorrgb#1{\color{black}}\def\colorgray#1{\color[gray]{#1}}\expandafter\def\csname LTw\endcsname{\color{white}}\expandafter\def\csname LTb\endcsname{\color{black}}\expandafter\def\csname LTa\endcsname{\color{black}}\expandafter\def\csname LT0\endcsname{\color{black}}\expandafter\def\csname LT1\endcsname{\color{black}}\expandafter\def\csname LT2\endcsname{\color{black}}\expandafter\def\csname LT3\endcsname{\color{black}}\expandafter\def\csname LT4\endcsname{\color{black}}\expandafter\def\csname LT5\endcsname{\color{black}}\expandafter\def\csname LT6\endcsname{\color{black}}\expandafter\def\csname LT7\endcsname{\color{black}}\expandafter\def\csname LT8\endcsname{\color{black}}\fi
  \fi
    \setlength{\unitlength}{0.0500bp}\ifx\gptboxheight\undefined \newlength{\gptboxheight}\newlength{\gptboxwidth}\newsavebox{\gptboxtext}\fi \setlength{\fboxrule}{0.5pt}\setlength{\fboxsep}{1pt}\begin{picture}(3400.00,3400.00)\gplgaddtomacro\gplbacktext{\csname LTb\endcsname \put(1530,1835){\makebox(0,0)[l]{\strut{}$\textbf{d}$}}}\gplgaddtomacro\gplfronttext{}\gplbacktext
    \put(0,0){\includegraphics{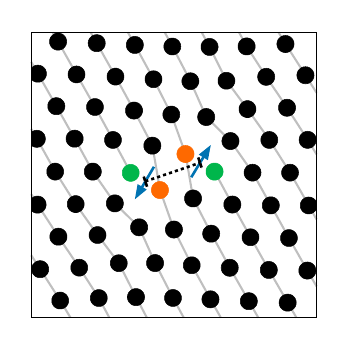}}\gplfronttext
  \end{picture}\endgroup
 }}
    \caption{Vacancy and conjugate dislocation pair in a triangular lattice. The particles in green has five neighbours and in orange has seven. The arrows represent the burgers vector of each dislocation.}
    \label{dislocpair}
\end{figure}

Let us consider that the density associated to the  Burguer's vector of a dislocation at position $\textbf{R}$ of the discrete lattice is given by:
\begin{equation}
\begin{split}
\label{eq:dislpair}
\textbf{b}_d(\textbf{r},\textbf{R})=\textbf{e}~\delta_{\textbf{r},\textbf{R}}.
\end{split}
\end{equation}
where $\delta_{\textbf{r},\textbf{R}}$ represent the standard Kronecker delta symbol and $\textbf{e}$ represent the Burguer's vector of the corresponding dislocation. Now we would like to extend the field defined in eq.~(\ref{eq:dislpair}) to the continuum to proceed with the calculation of the energy of a dislocation. Let us define this field as $\textbf{b}_c(\textbf{r})$. From dimensional arguments, it is natural to propose the following relation between both formulations:
\begin{equation}
    \int_{uc(\textbf{R})}\textbf{b}_c(\textbf{r}'-\textbf{R})d^2\textbf{r}'=A_{uc} \textbf{b}_d(\textbf{R},\textbf{R}),
\end{equation}
where $A_{uc}$ is taken as $\frac{1}{2}\sqrt{3}a^2$, considering that the dislocations "center" have to be positioned on the underlying triangular lattice of the solid. The previous relation lead us to the conclusion that $\textbf{b}_c(\textbf{r})=\frac{1}{2}\sqrt{3}a^2\textbf{e}\delta(\textbf{r})$. In this way the field corresponding to a conjugate dislocation pair --- as the one shown in Fig. \ref{dislocpair} (b) --- separated by a vector $\textbf{d}$ and oriented in the direction $\textbf{e}$ can be written as 
\begin{equation}
\begin{split}
\textbf{b}_c(\textbf{r})=\frac{\sqrt{3}a^2}{2}\left[\delta\left(\textbf{r}-\textbf{d}/2\right)-\delta\left(\textbf{r}+\textbf{d}/2\right)\right] \textbf{e}.
\end{split}
\end{equation}

The modulus of the vector $\textbf{e}$ is take it as one, which corresponds to the minimum possible value of the Burguer's vector of a dislocation in units of the lattice spacing. Now we can write our dislocation field in Fourier space as:
\begin{equation}
\begin{split}
\hat{\textbf{b}}(\textbf{q})=\frac{\sqrt{3}a^2}{2} (-2 i) \sin\left(\frac{\textbf q\cdot \textbf d}{2}\right) \textbf{e},
\end{split}
\end{equation}

It is not difficult to realize now that the energy of a dislocation pair will depend on the orientation of the vectors $\textbf{d}$, which is parameterized by the angle $\alpha$, and $\textbf{e}$, parameterized by the angle $\theta$, while both angles are measured respect to the $x$-axis. Consequently, we observe that the $x$ and $y$ components of the dislocation density field are given by:
\begin{equation}
\begin{split}
\hat{b}_x(\textbf{q})&=\frac{\sqrt{3}a^2}{2} (-2 i) \cos(\theta) \sin\left(\frac{\textbf q\cdot \textbf d}{2}\right),\\
\hat{b}_y(\textbf{q})&=\frac{\sqrt{3}a^2}{2} (-2 i) \sin(\theta) \sin\left(\frac{\textbf q\cdot \textbf d}{2}\right).
\end{split}
\end{equation}
In this way, the energy of a single dislocation can be calculated as one half of the total energy of a pair of conjugate dislocations, wich in units of $k_BT$ can be written as:
\begin{align}
\label{coreenergy}
\nonumber
\frac{E_c}{k_BT}&=\frac{K}{16\pi}\frac{1}{(A_{uc})^2}\int_{BZ} \frac{d^2q}{(2\pi)^2}\frac{4\pi}{q^2}\Big[\hat b_x(\textbf{q})\hat b_x(-\textbf{q})\left(1-\frac{q_x^2}{q^2}\right)\\
&+2\hat b_x(\textbf{q})\hat b_y(-\textbf{q})\left(-\frac{q_x q_y}{q^2}\right)+
\hat b_y(\textbf{q})\hat b_y(-\textbf{q})\left(1-\frac{q_y^2}{q^2}\right)\Big],
\end{align}
which after some simplifications can be recast in the following form
\begin{equation}
\begin{split}
    \frac{E_c}{k_BT}&=\frac{K}{16\pi}\frac{1}{A_{uc}^2}\int_{BZ} \frac{d^2q}{(2\pi)^2}\frac{12\pi a^4(\textbf{q}_\perp\cdot\textbf{e})^2}{q^4} \sin^2(\textbf{q}\cdot\textbf{d}/2),\\
    &=K\int_{BZ} \frac{d^2q}{(2\pi)^2}\frac{(\textbf{q}_\perp\cdot\textbf{e})^2}{q^4} \sin^2(\textbf{q}\cdot\textbf{d}/2).
\end{split}
\end{equation}
\indent To finally determine the actual value of the energy of a dislocation, we take the arithmetic average over all possible configurations in $\theta$ and $\alpha$, considering the do not have in fact one single type of defect but a continuous class. This procedure allow us to conclude that:
\begin{equation}
\begin{split}
    \left<(\textbf{q}_\perp\cdot\textbf{e})^2\right>_\theta&=\frac{q^2}{2},\\
    \left<\sin^2(\textbf{q}\cdot\textbf{d}/2)\right>_\alpha&=\frac{1}{2}\left(1-J_0(q d)\right),
\end{split}
\end{equation}
which lead us to the conclusion:
\begin{equation}\label{energy_core}
\begin{split}
    \frac{E_c}{k_B T}=K\int_{BZ} \frac{d^2q}{(2\pi)^2}\frac{1-J_0(qd)}{4 q^2}.
\end{split}
\end{equation}
\\
In order to reach a final mean value for the energy of a dislocation, we need as an input the distance between the dislocations in the pair, $d$. We can consider this distance as the minimum distance for which a dislocation pair is stable, since such pairs are the most relevant.

To understand better how to estimate such a quantity we performed M.D. simulations at $k_B T\sim 0$ (see main text) for the specific model under consideration (GCM). We used the ordered triangular lattice as initial condition for our simulation, removed or added a single particle and study the stationary configuration of the system given those specific initial conditions. Here is important to remark that vacancy or interstitial defects can be both considered as dislocation pairs, nonetheless we often regard them as different kind of defects since they have a higher symmetry \cite{Fisher1979} than an arbitrary dislocation pair. In our simulations we have noticed that vacancies are stable for $\rho\leq 0.3$, and naturally small fluctuations continuously deform its structure into pairs of dislocations with distance $d\sim \sqrt{3}a$ for $\rho>0.3$ -- both cases are represented in Fig. \ref{dislocpair} (b).

For higher temperatures, we have also noticed the presence of dislocation pairs more tightly bounded. Different from the defects found by relaxing the system from a vacancy initial condition, these pairs are not stable. They can be thus thought as ``virtual'' dislocation pairs, as they vanish quickly if we let the system relax at $k_B T\sim 0$. As an example of this line of reasoning, in ref. \cite{Fisher1979}, defects are considered as dislocation pairs only if the separation is greater than a certain threshold value necessarily higher than the lattice spacing of the underlying lattice.

Finally, if we consider $d=\sqrt{3}a$ in Eq. (\ref{energy_core}) and perform the numeric integration over the Brillouin zone, we can conclude our estimate for the energy of a dislocation, $E_c/k_BT=b K$, with $b\simeq 0.072$.

\end{document}